\documentclass[12pt]{aastex6}

\usepackage{graphicx,amssymb,amsmath,times}
\usepackage{longtable}
\usepackage{bm} 
\usepackage{url}
\usepackage[varg]{txfonts}
\usepackage[T1]{fontenc}
\usepackage[latin9]{inputenc}
\usepackage{natbib}
\def\simlt{\lower.5ex\hbox{\ltsima}}
\def\simgt{\lower.5ex\hbox{\gtsima}}

\def\farcm{\hbox{$\mkern-4mu^\prime$}}

\def\farcs{\hbox{$^{\prime\prime}$}~}

\def\kpc{{\rm\,kpc}}

\def\gtsim{\;\lower.6ex\hbox{$\sim$}\kern-6.7pt\raise.4ex\hbox{$>$}\;}
\def\ltsim{\;\lower.6ex\hbox{$\sim$}\kern-6.9pt\raise.4ex\hbox{$<$}\;}

\def\Sec{${}^{\prime\prime}$\llap{.}}
\def\deg{${}^\circ$}
\def\min{${}^{\prime}$}
\def\sec{${}^{\prime\prime}$}

\def\bmv{\hbox{\it B--V\/}}

\def\bmi{\hbox{\it B--I\/}}
\def\vmi{\hbox{\it V--I\/}}

\def\jmk{\hbox{\it J--K\/}}

\def\ngc#1{\hbox{NGC$\,$#1}}

\shorttitle{$\omega$ Centauri Optical {\it UBVRI\/} Photometry}
\shortauthors{Braga et~al.}

\begin{document}

\title{On the RR Lyrae stars in globulars: IV. $\omega$ Centauri Optical {\it UBVRI\/} 
Photometry\textsuperscript{*}}

\footnotetext[*]{Based in part on proprietary data and on data obtained from the
ESO Science Archive Facility under multiple requests by the authors; and in part
upon data distributed by the NOAO Science Archive. NOAO is operated by the
Association of Universities for Research in Astronomy (AURA) under cooperative
agreement with the National Science Foundation. This research also benefited
from the Digitized Sky Survey service provided by the Canadian Astronomy Data
Centre operated by the National Research Council of Canada with the support
of the Canadian Space Agency. A detailed description of the log of the
observations used in this investigation is given in Table~\ref{tbl:omegalog_opt}.\newline}

\author{V.~F.~Braga\altaffilmark{1,2}, P.~B.~Stetson\altaffilmark{3},
G.~Bono\altaffilmark{1,4},
M.~Dall'Ora\altaffilmark{5}, I.~Ferraro\altaffilmark{4}, G.~Fiorentino\altaffilmark{6},
L.~M.~Freyhammer\altaffilmark{7}, G.~Iannicola\altaffilmark{4}, M.~Marengo\altaffilmark{8}, 
J.~Neeley\altaffilmark{8}, E.~Valenti\altaffilmark{9},
R.~Buonanno\altaffilmark{1,10}, A.~Calamida\altaffilmark{11}, M. Castellani\altaffilmark{4}, 
R.~da Silva\altaffilmark{4,2}, S.~Degl'Innocenti\altaffilmark{12,13},
A.~Di Cecco\altaffilmark{10},
M.~Fabrizio\altaffilmark{10,2},
W.~L.~Freedman\altaffilmark{14},
G.~Giuffrida\altaffilmark{4,2},
J.~Lub\altaffilmark{15},
B.~F.~Madore\altaffilmark{16},
M.~Marconi\altaffilmark{5},
S.~Marinoni\altaffilmark{4,2},
N.~Matsunaga\altaffilmark{17},
M.~Monelli\altaffilmark{18},
S.~E.~Persson\altaffilmark{16},
A.~M.~Piersimoni\altaffilmark{10},
A.~Pietrinferni\altaffilmark{10},
P.~Prada-Moroni\altaffilmark{12,13},
L.~Pulone\altaffilmark{4},
R.~Stellingwerf\altaffilmark{19},
E.~Tognelli\altaffilmark{12,13},
A.~R.~Walker\altaffilmark{20}
}

\altaffiltext{1}{Department of Physics, Universit\`a di Roma Tor Vergata, via della Ricerca Scientifica 1, 00133 Roma, Italy}
\altaffiltext{2}{ASDC, via del Politecnico snc, 00133 Roma, Italy}
\altaffiltext{3}{NRC-Herzberg, Dominion Astrophysical Observatory, 5071 West Saanich Road, Victoria BC V9E 2E7, Canada}
\altaffiltext{4}{INAF-Osservatorio Astronomico di Roma, via Frascati 33, 00040 Monte Porzio Catone, Italy}
\altaffiltext{5}{INAF-Osservatorio Astronomico di Capodimonte, Salita Moiariello 16, 80131 Napoli, Italy}
\altaffiltext{6}{INAF-Osservatorio Astronomico di Bologna, Via Ranzani 1, 40127 Bologna, Italy}
\altaffiltext{7}{Jeremiah Horrocks Institute of Astrophysics, University of Central Lancashire, Preston PR1 2HE, UK}
\altaffiltext{8}{Department of Physics and Astronomy, Iowa State University, Ames, IA 50011, USA}
\altaffiltext{9}{European Southern Observatory, Karl-Schwarzschild-Str. 2, 85748 Garching bei Munchen, Germany}
\altaffiltext{10}{INAF-Osservatorio Astronomico di Teramo, Via Mentore Maggini snc, Loc. Collurania, 64100 Teramo, Italy}
\altaffiltext{11}{National Optical Astronomy Observatory, 950 N Cherry Avenue, Tucson, AZ 85719, USA}
\altaffiltext{12}{INFN, Sezione di Pisa, Largo Pontecorvo 3, 56127, Pisa, Italy}
\altaffiltext{13}{Dipartimento di Fisica ``Enrico Fermi'', Universit\`a di Pisa, Largo Pontecorvo 3, 56127, Pisa, Italy}
\altaffiltext{14}{Department of Astronomy \& Astrophysics, University of Chicago, 5640 South Ellis Avenue, Chicago, IL 60637, USA}
\altaffiltext{15}{Sterrewacht Leiden, Leiden University, PO Box 9513, 2300 RA Leiden, The Netherlands}
\altaffiltext{16}{The Observatories of the Carnegie Institution for Science, 813 Santa Barbara St., Pasadena, CA 91101, USA}
\altaffiltext{17}{Kiso Observatory, Institute of Astronomy, School of Science, The University of Tokyo, 
10762-30, Mitake, Kiso-machi, Kiso-gun, 3 Nagano 97-0101, Japan}
\altaffiltext{18}{Instituto de Astrof\'isica de Canarias, Calle Via Lactea s/n, E38205 La Laguna, Tenerife, Spain}
\altaffiltext{19}{Stellingwerf Consulting, 11033 Mathis Mtn Rd SE, Huntsville, AL 35803, USA}
\altaffiltext{20}{Cerro Tololo Inter-American Observatory, National Optical Astronomy Observatory, Casilla 603, La Serena, Chile}

\date{\centering Submitted \today\ / Received / Accepted }

\begin{abstract}
New accurate and homogeneous optical {\it UBVRI} photometry has been obtained for
variable stars in the Galactic globular $\omega$ Cen (\ngc{5139}).
We secured 8202 CCD images covering a time interval of 24 years and a sky area of
84$\times$48 arcmin. The current data were complemented with data available in the literature 
and provided new, homogeneous pulsation parameters (mean magnitudes, luminosity amplitudes,
periods) for 187 candidate $\omega$ Cen RR Lyrae (RRLs).
Among them we have 101~RRc (first overtone), 85~RRab (fundamental) and a 
single candidate RRd (double-mode) variables.
Candidate Blazhko RRLs show periods and colors that are intermediate between
RRc and RRab variables, suggesting that they are transitional objects.
The comparison of the period distribution and of the Bailey diagram indicates
that RRLs in $\omega$ Cen show a long-period tail not present in typical 
Oosterhoff II (OoII) globulars. The RRLs in dwarf spheroidals
and in ultra faint dwarfs have properties between Oosterhoff intermediate
and OoII clusters. Metallicity plays a key role in shaping the above
evidence. These findings do not support the hypothesis that
$\omega$ Cen is the core remnant of a spoiled dwarf galaxy.
Using optical Period-Wesenheit relations that are reddening-free
and minimally dependent on metallicity we find a mean distance
to $\omega$ Cen of 13.71$\pm$0.08$\pm$0.01 mag (semi-empirical 
and theoretical calibrations).
Finally, we invert the {\it I\/}-band Period-Luminosity-Metallicity relation to
estimate individual RRLs metal abundances. The metallicity distribution
agrees quite well with spectroscopic and photometric metallicity estimates
available in the literature.
\end{abstract}


\maketitle

\section{Introduction} \label{chapt_intro_omega}

The Galactic stellar system $\omega$ Cen lies at the crossroads of several open 
astrophysical problems. It is the most massive Milky Way globular cluster 
($4.05 \cdot 10^6 M_\odot [d/(5.5 \pm 0.2 \kpc)]^3$ where d is the distance,
\citealp{dsouza_e_rix2013}) and was the first to show 
a clear and well defined spread in metal-abundance 
\citep{norrisdacosta1995,johnson_e_pilachowski2010} in $\alpha$ and 
in s- and r-process elements \citep{johnson2009}. On the basis of the above peculiarities 
it has also been suggested that $\omega$ Cen and a few other massive Galactic 
Globular Clusters (GGCs) might have been the cores of pristine dwarf 
galaxies \citep{dacosta_e_coleman2008,marconi2014}.  

The distance to $\omega$ Cen has been estimated using primary and 
geometrical distance indicators.
The Tip of the Red Giant Branch (TRGB) was adopted by \citet{bellazzini2004,bono2008b}
with distances ranging from 13.65 to 13.70 mag. 
The {\it K\/}-band Period-Luminosity (PL) relations of RR Lyrae stars (RRLs) have been adopted by 
\citet{longmore1990,sollima2006b,bono2008b}.
The distance moduli they estimated range from 13.61 to 13.75 mag. On the other hand, 
$\omega$ Cen distance moduli based on the relations between luminosity 
and iron abundance for RRLs range from 
13.62 to 13.72 mag \citep{delprincipe_etal2006}.
The difference in distance between the different 
methods is mainly due to the intrinsic spread in the adopted diagnostics 
and in the reddening correction.   

Optical PL relations for SX Phoenicis stars were adopted by 
\citet{mcnamara2011} who found a distance of 13.62$\pm$0.05 mag.  
One eclipsing variable has been studied by \citet{kaluzny_etal2007},
and they found a distance modulus of 13.49$\pm$0.14 mag and 
13.51$\pm$0.12 mag for the two components. The key advantage in dealing with 
eclipsing binaries is that they provide very accurate geometrical distances 
\citep{pietrzynski_etal2013}.
Estimates based on cluster proper motions provide distance estimates that are 
systematically smaller than obtained from the other most popular distance 
indicators (13.27 mag, \citet{vanleeuwen_etal2000};  13.31$\pm$0.04 mag, 
\citet{watkins_etal2013}). The reasons for this difference are not clear yet.   

The modest distance and the large mass of $\omega$ Cen make this stellar system 
a fundamental  laboratory to constrain evolutionary and pulsation properties 
of old (t>10 Gyr) low-mass stars. The key advantage in dealing 
with stellar populations in this stellar system is that they cover a broad 
range in metallicity 
(--2.0$\lesssim$ [Fe/H] $\lesssim$--0.5, \citet{pancino2002};
--2.5$\lesssim$ [Fe/H] $\lesssim$+0.5, \citet{calamida_etal2009};
--2.2$\lesssim$ [Fe/H] $\lesssim$--0.6, \citet{johnson_e_pilachowski2010})   
and they are located at the same distance \citep{castellani_etal2007}. 
Moreover, the high total stellar mass does provide 
the opportunity to trace fast evolutionary phases 
\citep{monelli_etal2005,calamida_etal2008} together with exotic 
\citep{randall_etal2011} and/or compact objects \citep{bono_etal2003}.   

Exactly for the same reasons mentioned before $\omega$ Cen was a crucial 
crossroads for RRLs.   The first detailed investigation of RRLs 
was provided more than one century ago in a seminal investigation by 
\citet{bailey1902}. Using a large set of photographic plates he identified 
and characterized by eye 128 RRLs, providing periods, amplitudes 
and a detailed investigation of the shapes of the light curves. In 
particular, he suggested the presence of three different kind of pulsating variables
(RRa, RRb, RRc) in which the luminosity variation amplitude steadily decreases 
and the shape of the light curve changes from sawtooth to sinusoidal.   
This investigation was supplemented more than thirty years later by 
\citet{martin1938} on the basis of more than 400 photographic plates collected  by 
H. van Gent on a time interval of almost four years and measured with a 
microdensitomer. He provided homogeneous photometry and very accurate 
periods for 136 RRL variables.  

We needed to wait another half century to have a detailed and almost complete 
census of RRL in $\omega$  Cen based on CCD photometry, by the OGLE 
project \citep{kaluzny_etal1997,kaluzny_etal2004}. 
They collected a large number of
CCD images in {\it V\/} and {\it B\/} covering a time 
interval of  three years \citep{kaluzny_etal1997} and one and half years
\citep{kaluzny_etal2004} and provided a detailed 
analysis of the occurrence of the Blazhko effect \citep[a modulation of
the light amplitude on time scales from tens of days to years,][]{blazhko1907}. 
A similar analysis was 
also performed by \citet{weldrake_etal2007} using the observing facility 
and photometric system of the MACHO project. They collected 875 optical 
images covering a period of 25 days.  

A detailed near-infrared (NIR) analysis was performed by 
\citet{delprincipe_etal2006} 
using time series data collected with SOFI at NTT. 
They provided homogeneous {\it JK$_s$\/} photometry for 180 variables and provided 
a new estimate of the $\omega$ Cen distance modulus using the {\it K\/}-band PL 
relation (13.77$\pm$0.07 mag). A similar analysis was recently performed 
by \citet{navarrete_etal2015} based on a large set of images 
collected with the VISTA telescope. 
They provided homogeneous {\it JK$_s$\/} photometry for 189 probable member RRLs
(101 RRc, 88 RRab) and discussed the pulsation properties 
of the entire sample in the NIR. In particular, they provided new NIR
reference lines for Oosterhoff I (OoI) and Oosterhoff II (OoII) clusters. Moreover, they further
supported the evidence that RRab in $\omega$ Cen display properties
similar to OoII systems. These investigations have been complemented with
a detailed optical investigation covering a sky area of more than 50 square
degrees by \citet{fernandeztrincado2015}. They detected 48 RRLs and the
bulk of them (38) are located outside the tidal radius. However, detailed
simulations of the different Galactic components and radial velocities for
a sub-sample of RRLs indicate a lack of tidal debris around the cluster.

This is the fourth paper of a series focussed on homogeneous optical, near-infrared,
and mid-infrared photometry of cluster RRLs. The structure of the paper is as 
follows. In \S~2 we present the optical multi-band {\it UBVRI\/} photometry that we 
collected for this experiment together with the approach adopted to perform 
the photometry on individual images and on the entire dataset. In subsection 3.1 
we discuss in detail the identification of RRLs and the photometry we collected from the 
literature to provide homogeneous estimates of the RRL pulsation parameters.    
Subsection 3.2 deals with the period distribution, while subsection 3.3 
discusses the light curves and the approach we adopted to estimate the mean 
magnitudes and the luminosity variation amplitudes. 
The Bailey diagram (luminosity variation amplitude vs 
period) is discussed in \S~3.4, while the amplitude ratios are considered in \S~3.5. 
Section 4 is focussed on the distribution of RRLs in the color-magnitude 
diagram (CMD) and on the topology of the instability strip. In \S~5 we perform a detailed 
comparison of the period distribution and the Bailey diagram of $\omega$ Cen 
RRLs with the similar distributions in nearby gas-poor systems (globulars, 
dwarf galaxies). Section 6 deals with RRL diagnostics, namely the PL and the Period-Wesenheit 
(PW) relation, while in \S~7 we discuss the new distance determinations to 
$\omega$ Cen  based on optical PW relations. Section 8 deals with the 
metallicity distribution of the RRLs, based on the {\it I}-band PL relation, and the 
comparison with photometric and spectroscopic estimates available 
in the literature. Finally, \S~9 gives a summary of the 
current results together with a few remarks concerning the future of this 
project. 

\section{Optical photometry}\label{chapt_obs_opt}

We provide new accurate and homogeneous calibrated 
multi-band {\it UBVRI\/} photometry for the candidate RRLs in $\omega$ Cen. 
The sky area covered by our calibrated photometry is roughly 
57 \farcm$\times$56 \farcm around the cluster center (see the end 
of this section). We acquired 8202 
optical CCD images of $\omega$ Cen from 
proprietary datasets (6211 images, 76\%) 
and public archives and extracted 
astrometric and photometric measurements from them using well established
techniques (see, e.g., \citealp{stetson2000,stetson2005}, and references therein).
Among these we were able to photometrically calibrate 7766 images
(including 320 {\it U\/}-, 2632 {\it B\/}-, 3588 {\it V\/}-, 339 {\it R\/}-, 
and 887 {\it I\/}-band images) covering a time 
interval of slightly over 24 years. Table~\ref{tbl:omegalog_opt} gives the log of observations 
and a detailed description of the different optical datasets adopted in this 
investigation. Note that the largest datasets are {\it danish95} 
(1786 CCD images)\footnote{This dataset also includes 140 CCD images that were collected 
in 1996. They were included in the {\it danish95} dataset due to the limited sample size.}  
and {\it danish98} (1981 CCD images).  The {\it danish99} dataset also includes 
a sizable number of exposures (632 CCD images), but they were collected on 
two nights separated by seven days. For this reason, the {\it danish99} dataset is 
very useful to have a guess of the shape of the light curve, but the period determinations 
based on this dataset in isolation are not as accurate as those based on datasets 
covering a larger time interval. The {\it B\/}-band photometry based on {\it danish95} and on  
{\it danish98} images is less accurate when compared with the other datasets. The 
{\it danish98} dataset showed large variations of the photometric zero-point 
with position on the chip. The large number of local standards allowed us to take 
account for this positional effect.   
The photometry based on all the other datasets 
was labeled {\it other\/} and provides most of the time interval 
covered by our photometric catalog. Note that these data were collected with 
several ground-based telescopes available at CTIO (0.9m, 1.5m, Blanco 4m), 
ESO (0.9m, MPI/ESO 2.2m, NTT, VLT), and SAAO (1m). 
In passing we also note that the current dataset was also built up to detect 
fast evolving objects, i.e., objects experiencing evolutionary changes on 
relatively short time scales.

The defining of local standards in the field of $\omega$ Cen was performed following  
the same criteria discussed in our previous work on M4 \citep[free from blending, 
a minimum of three observations, standard error lower than 0.04~mag and intrinsic 
variability smaller than 0.05~mag,][]{stetson2014}. As a whole 4,180 stars satisfy 
these requirements and 4,112 of these have high-quality photometry (at least five 
observations, standard error $<$ 0.02~mag and intrinsic variability smaller 
than 0.05~mag) in at least two bands:  3,462 in {\it U\/}, 4,112 in {\it B\/}, 
4,106 in {\it V\/}, 875 in {\it R\/}, and 3,445 in {\it I\/}. 

These stars have been used as a local reference for the photometric
calibration of 847,138 stars in the field of $\omega$ Cen from the final 
ALLFRAME reduction \citep{stetson1994}. 
The median seeing of the different datasets is 1.2\sec, but our 25-th percentile 
is 0.86\sec, and the 10-th percentile is 0.65\sec.
We measured stars in up to 2,000 images covering the innermost cluster regions
(13.4\min$\times$13.5\min); 
this means that the stars located there, were observed in $\sim$200 images with 
seeing better than 0.65\sec. The cluster regions in which we measured stars in 
up to 100 images is 36.6\min$\times$31.6\min; this means that roughly ten images 
were collected with a seeing better than 0.65\sec. The use of ALLFRAME means 
that detections visible in these images and their positions were also used to 
fit those same stars in the poorer-seeing images.  
This analysis resulted in 583,669 stars with calibrated photometry in all
three of {\it B\/}, {\it V\/}, and {\it I\/}; 202,239 of them had calibrated
photometry in all five of {\it U\/}, {\it B\/}, {\it V\/}, {\it R\/}, and {\it I\/}. 
A more detailed analysis of the current multiband photometric dataset will be provided 
in a forthcoming paper (Braga et al. 2016, in preparation).   

The astrometry of our photometric catalog is on the system of the USNO A2.0 catalog
\citep{monet1998}. 
The astrometric accuracy is 0.1\sec and allows us to provide 
a very accurate estimate of $\omega$ Cen's centroid. Following the approach 
applied to Fornax by \citet{stetson1998} we found $\alpha_{center}$
= 13$^h$ 26$^m$ 46.71$^s$, $\delta_{center}$ = --47\deg 28 \farcm 59\Sec9. 
The stars adopted to estimate the position of the cluster center have 
{\it V\/}-band magnitude between 16 and 20 and are located (with varying 
weights) up to $\sim$14~\farcm ~from the adopted center.
The mean epoch associated to these coordinates is May 2004.
New and accurate structural parameters will be provided in a forthcoming paper.

\section{RR Lyrae stars}

\subsection{Identification}\label{chapt_rrind_omega}

We adopted the online reference catalog for cluster variables 
``Catalogue of Variable Stars in Globular Clusters'' by C.~Clement 
\citep[][updated 2015]{clement2001}. This catalog lists 456 variables 
(197 candidate RRLs) within the truncation radius ($r_t$=57.03 arcmin, \citealp{harris1996}) 
of $\omega$ Cen and it is mostly 
based on the detailed investigations of \citet{kaluzny_etal2004}. This catalog 
was supplemented with more recent discoveries by \citet{weldrake_etal2007} 
and \citet{navarrete_etal2015}. Two candidate field RRLs 
NV457 and NV458 discovered by \citet{navarrete_etal2015} were not 
included in the online catalog. To provide accurate and homogeneous 
photometry for the entire sample of RRLs along the line of sight 
of $\omega$ Cen, they were included in the current sample.
We have also removed the field star V180 from the sample, following \citet{navarrete_etal2015}
that classify it as a W UMa binary star on the basis of its color and pulsation amplitude.
Finally, we have included V175, recently recognized as
a field RRL by \citet{fernandeztrincado2015}, that
updated the uncertain classification of this object by \citet{wilkens1965}.
We ended up with 199 candidate RRLs. 
Eight out of the 199 objects are, 
according to their mean magnitudes and proper motions, candidate 
field stars \citep{vanleeuwen_etal2000,navarrete_etal2015}.
Note that we also included two variables with periods similar to 
RRLs but for which the classification is not well established, 
namely NV366 \citep{kaluzny_etal2004} and the candidate field variable 
NV433 \citep{weldrake_etal2007}. 

To overcome possible observational biases in the current RRL sample, 
the identification of variable stars was performed {\em ab initio} 
using the Welch-Stetson \citep[WS,][]{welchstet,stetson1996} index.
We adopted our own photometric catalog and we identified 176 candidate 
RRLs they had a WS index larger than 1.1. This list was cross matched with 
the Clement's catalog and we found that all of them were already known.
We have then compared the individual coordinates based 
on our astrometric solution with the ones given in the literature and we found that 
the median difference of the coordinates is 0\Sec41, with a standard deviation from 
the median of 0\Sec26. The difference for the entire sample is smaller than 2\sec; 
only for six out of the 176 stars is the difference between 1\sec and 2\sec.

The above data were supplemented with unpublished Walraven {\it WULBV\/} photometry for two 
RRLs---V55 and V84---collected by J.~Lub in 1980-1981 at the Dutch telescope in 
La Silla. The Walraven photometry was transformed into the standard Johnson-Kron-Cousins
photometric system ({\it UBV\/}) using the 
transformations provided by \citet{brandwouterloot1988}.
Moreover, we supplemented the photometry of the two variables
observed by J. Lub with {\it UBV\/} photoelectric 
photometry from \citet{sturch1978}.
{\it The number of RRLs for which we provide 
new astrometry is 186, while those for which 
we provide new photometry is 178}.

For nine out of the remaining 21 stars, we were able to recover
optical photometry in the literature:
a)---V281 and V283 from OGLE \citep[{\it V\/} band,][]{kaluzny_etal1997};
b)---V80, V177, NV411 and NV433 \citep[{\it V+R\/} band,][]{weldrake_etal2007};
c)---V172, NV457 and NV458 from CATALINA 
\citep[{\it V\/} band,][]{drake_etal2009,drake13a,drake13b,drake14,torrealba15}.

We performed detailed tests concerning the 
photometric zero-point using the objects
in common and we found that there is no difference,
 within the photometric errors,
between the current photometry and the photometry provided by OGLE, CATALINA,
\citet{sturch1978} and by J. Lub.
On the other hand, we have not been able to 
transform into the standard photometric
system the photometry collected by \citet{weldrake_etal2007} 
for the four variables
V80, V177, NV411 and NV433. For these four 
objects we only provided a homogeneous
period determination, while the other five were fully characterized (period,
mean magnitude, amplitude).

The positions of $\omega$ Cen RRLs based on 
our astrometric solution are listed in
columns 2 and 3 of Table~\ref{tab:pos_rr}, together with their literature and
current pulsation period (columns 4 and 5 in Table~\ref{tab:pos_rr}).
The epoch of the mean magnitude along the rising branch \citep{inno2015}
and of the maximum light are listed in column 6 to 9 of
Table~\ref{tab:pos_rr} together with the photometric band adopted
for the measurements. Note that the above epochs have been estimated using
the spline fits of the light curves discussed in Section~\ref{par:omegacen_lightcurves}.

\begin{figure*}[t]
\centering
\figurenum{1}
\includegraphics[width=14cm]{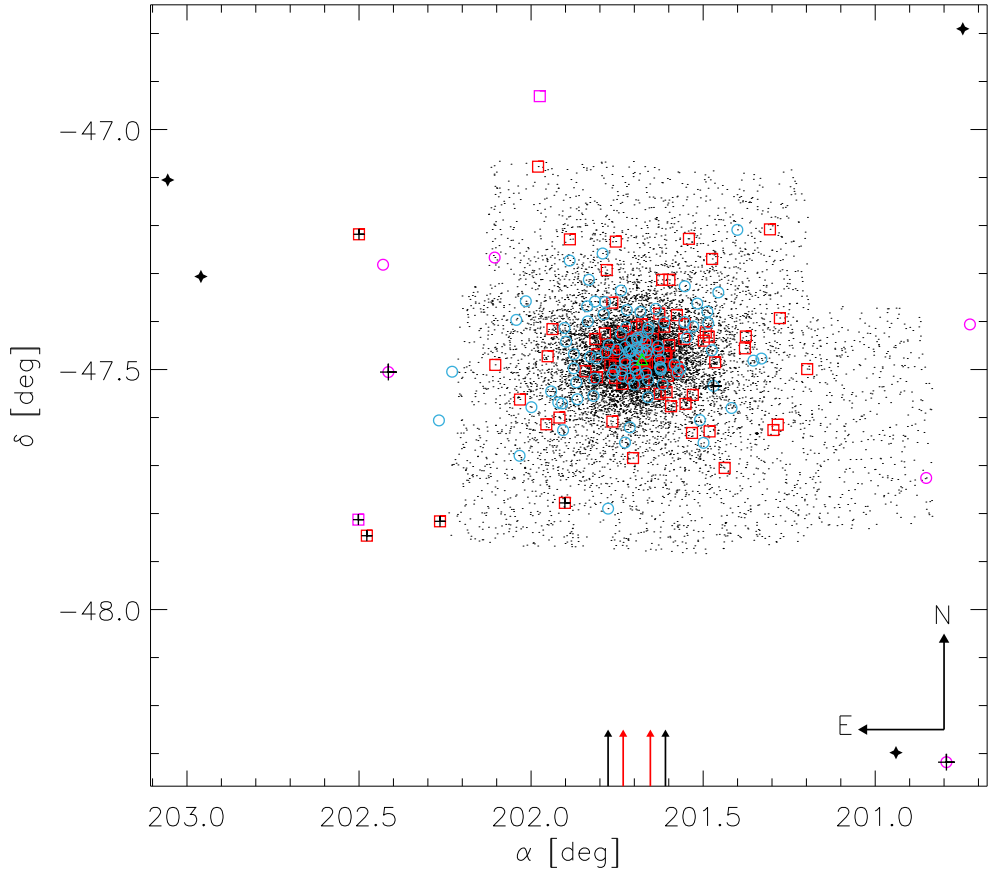}
\vspace*{1.0truecm}
\caption{Spatial distribution of the entire sample of $\omega$ Cen RRLs (199).  
The current optical photometry covers a sky area of $\approx$25\ \farcm across 
the cluster center. The upward red and black arrows plotted on the right ascension 
axis display the core and the half-mass radius \citep{harris1996}. The black arrows  
plotted in the bottom right corner display the orientation.
The squares and open circles display the position of RRab (90) and RRc (104)
variables, respectively. The candidate RRd variable (V142) is marked 
with a green triangle. Magenta squares (RRab) and circles (RRc)
mark the position of the eight variables for which we have retrieved
periods and either optical or NIR mean magnitudes in the literature
and for which a mode classification is possible. Black stars mark the 
position of the four variables for which we do not have solid pulsation
paramter estimates and mode classification. The pluses mark the position of 
the eight candidate field RRLs.}
\label{fig:omegaradec}
\end{figure*}

The sky distribution of the 187 candidate RRLs for which we have estimated 
pulsation parameters is shown in Fig.~\ref{fig:omegaradec}, where red squares
and light blue circles mark the position of fundamental (RRab) and 
first-overtone (RRc) RRLs (see \S 3.2).  The candidate RRd variable V142 
(see notes on individual variables in the Appendix) is marked with a green 
triangle.

We retrieved mean optical magnitude and periods from the literature 
for three variables: V151 \citep{martin1938}, V159 \citep{vangent1948} and for 
V175 \citep{fernandeztrincado2015}. Mean NIR magnitudes and periods 
for five variables (V173, V181, V183, V455, V456) were retrieved from
\citet{navarrete_etal2015}. Magenta squares (RRab) and circles (RRc)
mark the position of these eight variables. The four candidate variables 
identified by \citet[][V171, V178, V179]{wilkens1965} and by 
\citet[][V182]{sawyerhogg1973} for which we do not have solid estimates
of the pulsation parameters and mode classification are marked with black stars.   
Among the 195 RRLs for which the pulsation characterization has been 
performed we have 104 RRc, 90 RRab and a single RRd variable.   

We performed a number of statistical tests concerning the radial 
distribution of RRab and RRc, but no clear difference was found.

\subsection{Period distribution}\label{par:perioddistribution}

To take full advantage of the observing strategy adopted to collect 
the time series we used two independent methods to determine the periods: 
the string method \citep{stetson1996,stetson1998a} and our variant of the 
Lomb-Scargle (LS) method \citep{scargle82}. 
The key advantages of these methods are: 
a) they use multi-band photometry simultaneously; b) they take account for 
intrinsic photometric errors. We have checked that, within 0.002 days,
period estimates based on the two methods agree quite well with each 
other. The periods based on the Lomb-Scargle method also agree with 
those given in the Clement catalog. 
The difference between the Lomb-Scargle and the Clement
periods is typically smaller than 0.0001 days. Only 28 variables 
show a difference larger than 0.0001 days, but none has a difference 
larger than 0.001 days. 
Table~\ref{tab:pos_rr} only gives the periods based on the Lomb-Scargle 
method, because this method was also used for the variables with 
photometry only available in the literature. A preliminary analysis
on the uncertainties of the periods suggests us that they cannot be larger than
$1\cdot10^{-6}$ days.

The period derivatives of RRLs in $\omega$ Cen have been investigated 
by \citet{jurcsik2001}. They collected photometric data available in the 
literature covering more than one century. They found that a sizable 
sample of RRab display a steady increase in their period, thus 
supporting the redward evolution predicted by Horizontal Branch (HB) models \citep{bono2016}.
On the other hand, the RRc showed irregular trends in period changes. This 
indicates that period changes are affected by evolutionary effects and by other 
physical mechanisms that have not been fully constrained \citep{renzinisweigart1980}. 
We plan to provide more quantitative constraints of the period changes after the 
analysis of NIR images we have already collected, since they will allow us to 
further increase the time interval covered by our homogeneous photometry.        

It is well known that $\omega$ Cen hosts a sizable sample
of RRc with periods longer than 0.4 days \citep{kaluzny_etal2004}. 
To constrain the pulsation mode of the candidate RRLs,
we need to take account of their distribution in the Bailey diagram (period 
vs luminosity variation amplitude, see Section~\ref{par:bailey}). 

The current data allowed us to confirm the pulsation mode of the
current candidate RRLs; they are listed in the last column of
Table~\ref{tab:phot_rr_opt}. 
Using either optical or NIR mean magnitudes (see \S 3.1) as a 
selection criterion to discriminate between candidate field and 
cluster RRLs, we found that the candidate cluster RRLs number 187, 
and among them 101 are RRc, 85 are RRab and one single candidate 
RRd variable.

To make the separation between field and cluster stars more clear
the former in Fig.~\ref{fig:omegaradec} were marked with a plus sign. 
As expected, field candidates tend to be located between the half-mass 
radius \citep[$r_h =$ 5 arcmin,][]{harris1996}
and the tidal radius \citep[$r_t =$ 1.2 degrees,][]{marconi2014} of the cluster.

Note that, according to \citet{weldrake_etal2007} and to \citet{navarrete_etal2015}, 
the classification of the variable NV433, that has a peculiar 
light curve, is unclear. However, its apparent magnitude 
\citep[{\it K\/}$\sim$14.151 mag][]{navarrete_etal2015} seems to suggest that it is 
a candidate field variable.  

\begin{figure*}[t]
\centering
\includegraphics[width=10cm]{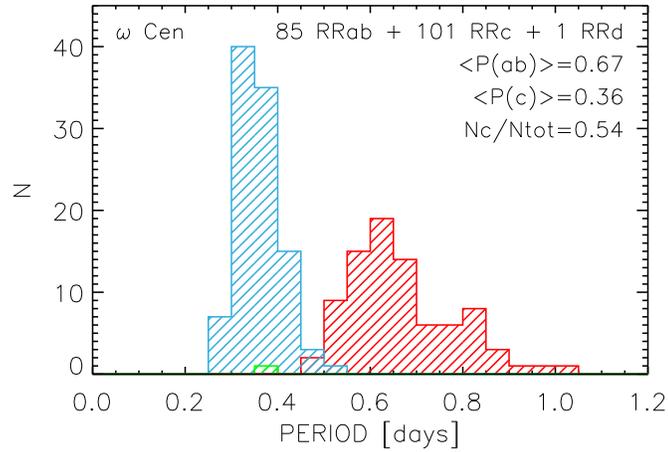}
\vspace*{1.0truecm}
\figurenum{2}
\caption{Period distribution for RRab (red) and RRc (light blue) RRLs in $\omega$ Cen. 
The candidate RRd variable (V142) is plotted in green.  
The number of RRab, RRc and RRd candidate RRLs are labeled 
together with the mean periods of RRab and RRc and the ratio between 
the number of RRc and the total number of RRLs 
($N_c$/$N_{ab}+N_d+N_c$). See text for more details.}
\label{fig:omegacen_histoperiod}
\end{figure*}

The period distribution plotted in Fig.~\ref{fig:omegacen_histoperiod} 
shows---as expected---a prominent peak for RRc (light blue shaded area) 
with roughly 20\% of the variables (18 out of 101) with periods longer 
than 0.4 days.

The RRab show a broad period distribution ranging from 
0.47 days to roughly one day (red shaded area). Long-period (P$\geq$0.82--0.85 days) 
RRLs are quite rare in Galactic globulars. Several of them have also been 
identified in two peculiar Bulge metal-rich globulars---\ngc{6388}, \ngc{6441} 
\citep{pritzl2001,pritzl2002}---and in the Galactic field \citep{wallerstein2009}.
Whether they are truly long-period RRLs or short-period Type II Cepheids (TIICs) 
is still a matter of lively debate \citep{soszynski2011,marconi2011}.  
In the current investigation we are assuming, following the OGLE team, that 
the transition between RRLs and TIICs takes place across one day.  
More quantitative constraints on this relevant issue will be addressed in 
a future paper.  

The ratio between the number of RRc and the total number of RRL 
($N_{tot}$=$N_{ab}$+$N_d$+$N_c$) is quite large 
($N_c$/$N_{tot} = 0.54$), roughly $\sim$0.1 larger than
the typical ratio of Oosterhoff II (OoII) clusters: $N_c$/$N_{tot}\sim0.44$, 
while the same ratio in Oosterhoff I (OoI) clusters is $N_c$/$N_{tot}\sim0.29$
\citet{oosterhoff1939,castellaniquarta1987,caputo1990}.
The mean Fundamental (F) period is $<P_{ab}> = 0.668$ days, i.e., quite similar to 
OoII clusters, since they have $<P_{ab}>\sim0.651$ days, while OoI 
clusters have $<P_{ab}>\sim0.557$ days. 
The mean First Overtone (FO) period is $<P_{c}>=0.359$ days, once again similar to 
OoII clusters, since they have $<P_{c}>=0.356$ days, while OoI clusters  
have $<P_{c}>=0.312$ days.
However, these mean parameters should be treated with caution, since they 
have been estimated using the same selection criteria adopted by 
\citet{fiorentino2015}, i.e., we only took into account GCs hosting at least 
35 RRLs.
A more detailed comparison with different Oosterhoff groups and with RRLs 
in nearby stellar systems is given in Section~\ref{chapt_comparison_oosterhoff}.

\subsection{Light curves}\label{par:omegacen_lightcurves}

The observing strategy of the large optical datasets adopted in this investigation 
was focussed on RRLs. The main aim was an extensive and homogeneous 
characterization of their pulsation properties (period, mean magnitudes, amplitudes 
and epochs of minimum and maximum light). The time coverage (24 years) and the 
approach adopted to perform simultaneous multiband photometry allow us to provide 
very accurate period determinations (see Section~\ref{par:perioddistribution}).

This experiment was also designed to provide accurate estimates of period 
variations, but this topic will be addressed in a forthcoming paper. 
This is the reason why we collected a few 
hundred phase points in a single band on individual nights. More importantly, 
we collected more than one thousand phase points on a time interval of 
one to two weeks. As a whole, this extremely dense sampling provides us 
very good phase coverage for both short and long-period RRLs. 
However, the phase coverage is marginally affected by alias in the 
transition between RRL and short-period TIIC (BL Herculis), 
i.e., in the period range across $\sim$1.0 day.

\begin{figure*}[t]
\centering
\includegraphics[width=17cm]{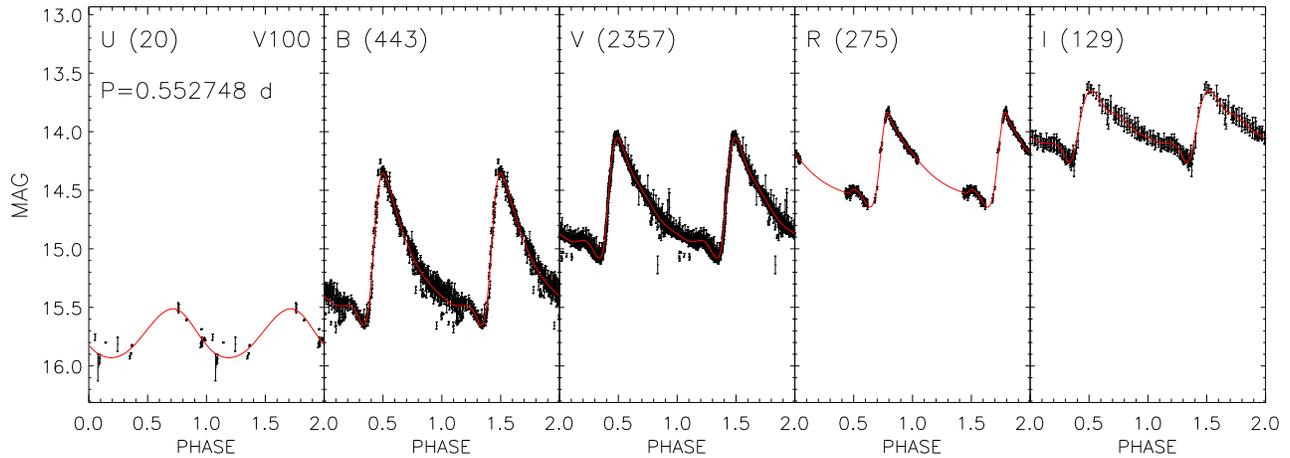}
\figurenum{3}
\caption{From left to right {\it UBVRI\/} light curves of the RRab variable V100. 
Red lines show the spline fits adopted to calculate mean magnitudes, 
luminosity variation amplitudes and the epochs of mean magnitude and maximum light. 
Data and analytical fits cover two full pulsation cycles. The number of phase 
points per band are displayed in parentheses. The period is also labeled. 
The vertical error bars display intrinsic errors of individual data points.}
\vspace*{0.7truecm}
\label{fig:omega_v100}
\end{figure*}

\begin{figure*}[t]
\centering
\figurenum{4}
\includegraphics[width=17cm]{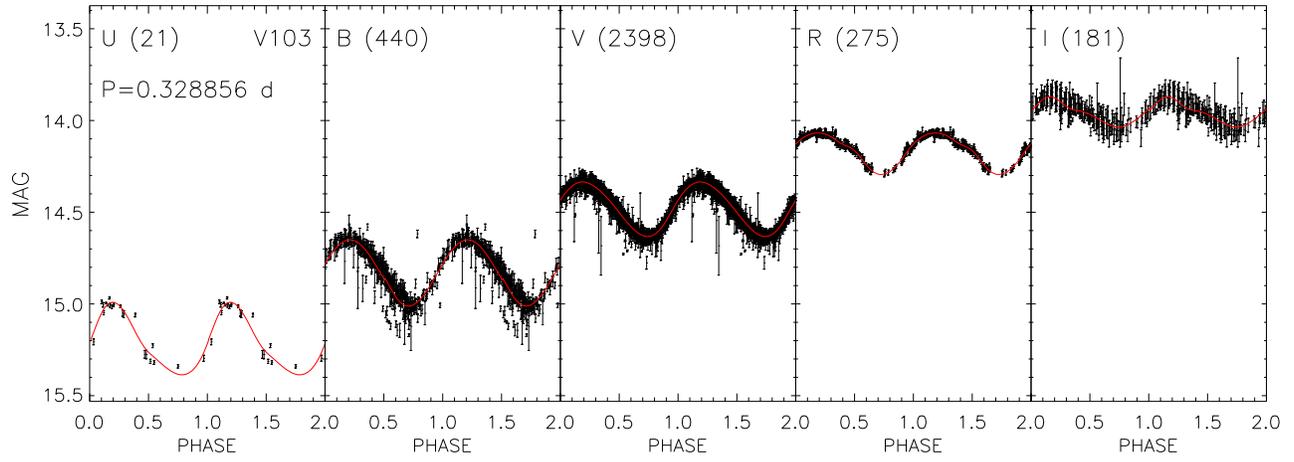}
\caption{Same as Fig.~\ref{fig:omega_v100}, but for the RRc variable V103.}
\vspace*{0.7truecm}
\label{fig:omega_v103}
\end{figure*}

\begin{figure*}[t]
\centering
\figurenum{5}
\includegraphics[width=17cm]{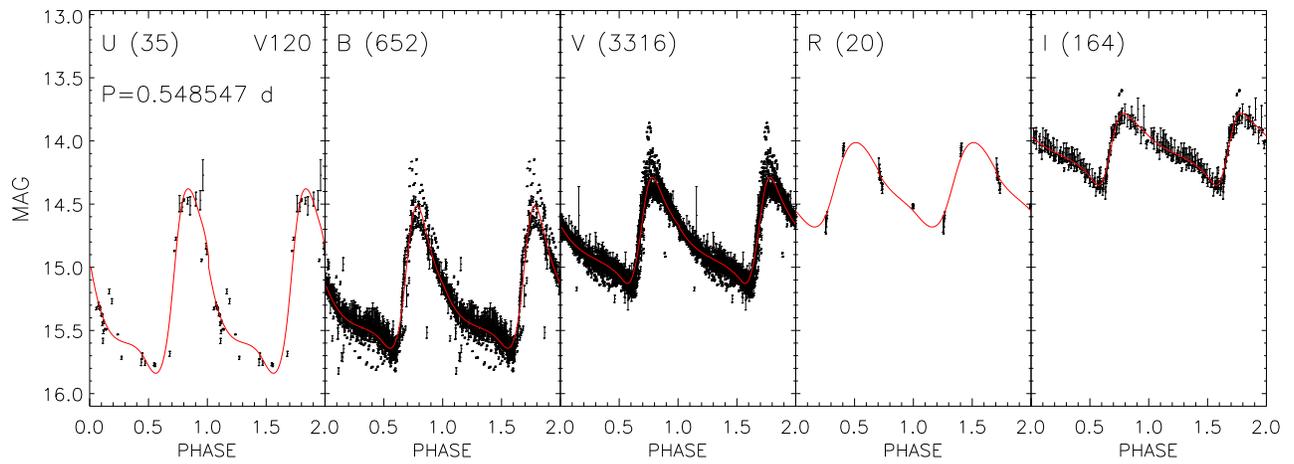}
\caption{Same as Fig.~\ref{fig:omega_v100}, but for the candidate Blazhko variable V120.} 
\label{fig:omega_blazhko}
\end{figure*}

The results of this observing strategy are visible 
in Figures~\ref{fig:omega_v100},~\ref{fig:omega_v103} 
and~\ref{fig:omega_blazhko} that show, from left to right, a selection of 
optical light curves in the {\it UBVRI\/} bands for a RRL star
pulsating in the F mode (V100), 
in the FO mode (V103) and a RRab 
variable affected by Blazhko (V120). The number 
of phase points per band and the period are also labelled. The 
vertical bars display individual photometric errors. They are 
of the order of $\sigma_U\sim$0.026, $\sigma_B\sim$0.025, 
$\sigma_V\sim$0.014, $\sigma_R\sim$0.012 and $\sigma_I\sim$0.035.

Solid red lines in Figures ~\ref{fig:omega_v100},~\ref{fig:omega_v103} 
and~\ref{fig:omega_blazhko} show  the spline fits that we adopted to 
derive mean magnitudes, amplitudes and epochs of mean and maximum light 
of RRLs. The {\it UBVRI\/} mean magnitudes of the candidate RRLs were 
derived by intensity-averaging the spline fits over a full pulsation 
cycle. They are listed in columns 2 to 6 of Table~\ref{tab:phot_rr_opt}. 
The column 12 of the same table gives the photometric quality index 
of the individual light curves in the different bands. It is zero for no phase coverage, one for 
poor phase coverage, two for decent coverage and three for good phase coverage.  
The errors of the mean magnitudes have been determined as the weighted 
standard deviation between the spline fit and the individual phase points. 
We found that the errors on average, for good quality light curves, are:     
$\sigma_B=0.02$ mag, $\sigma_V=0.01$ mag, 
$\sigma_R=0.01$ mag and $\sigma_I=0.03$ mag. The same errors for 
decent quality light curves are: 
$\sigma_U=0.02$ mag, $\sigma_B=0.02$ mag, $\sigma_V=0.01$ mag, 
$\sigma_R=0.02$ mag and $\sigma_I=0.04$ mag. 
The mean magnitudes of the objects for which the light curve coverage 
is poor, typically in the {\it U\/}-band, was estimated as the median of the 
measurements and their errors range from 0.04 ({\it I\/} band)
to 0.11 mag ({\it U\/} band).    
The luminosity variation amplitudes in the {\it UBVRI\/} bands of the candidate RRLs for 
which we have either our photometry or literature photometry have been 
estimated as the difference between the minimum and the maximum of the 
spline fit. They are listed in columns 7 to 11 of Table~\ref{tab:phot_rr_opt}.
Note that the {\it U\/}-band amplitudes are available only for a limited 
number of variables. Moreover, the minimum and maximum amplitudes 
of the candidate Blazhko RRLs and of the candidate 
RRd variable were estimated as the amplitudes of
the lower and the upper envelope of the observed data points.

\subsection{Bailey diagram}\label{par:bailey}

\begin{figure*}[t]
\centering
\figurenum{6}
\includegraphics[width=8cm]{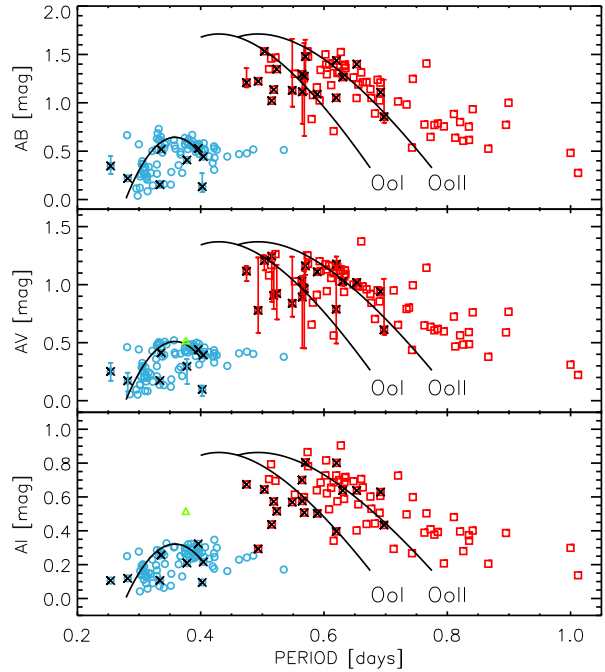}
\vspace*{1.0truecm}
\caption{Top: Bailey diagram, {\it B\/}-band amplitude versus period, for candidate 
$\omega$ Cen RRLs. RRab and RRc stars are plotted as red squares and light blue 
open circles, respectively. The candidate RRd variable (V142) is marked with a green 
triangle. {\it Its abscissa is fixed at the period of the main (FO) mode.} 
Candidate Blazhko RRLs are marked with a black cross. The solid black 
lines overplotted on the RRab stars display the loci typical of OoI and OoII GCs 
\citep{cacc2005}. The solid black lines overplotted on the RRab stars display the loci typical 
of OoI and OoII GCs \citep{cacc2005}. The solid black line overplotted on the 
RRc stars shows the locus of RRc typical of OoII GCs \citep{kunder2013c}. 
Note that the latter relation was originally provided by \citet{kunder2013c} 
for the {\it V\/}-band. It was transformed into the {\it B\/}-band using 
$AB/AV$=1.26. 
Middle: Same as the top, but for the {\it V\/}-band amplitude. The vertical 
red bars display the range in luminosity amplitude of candidate Blazhko RRLs. 
The Oosterhoff relations for RRab stars were originally provided by 
\citep{cacc2005} for the {\it B\/}-band. They were transformed into the 
{\it V\/}-band using $AB/AV$=1.25. 
Bottom: Same as the top, but for the {\it I\/}-band amplitude.
The original Oosterhoff relations for RRc and RRab stars were transformed 
into the {\it I\/}-band using $AI/AV$=0.63 (see \S~\ref{chapt_amplratio}).}
\label{fig:omegacenbailey}
\end{figure*}

The Bailey diagram---period vs luminosity variation amplitude---is a powerful diagnostic 
for variable stars, being reddening- and distance-independent \citep{smith2011}. 
Moreover, the luminosity variation amplitudes are also minimally affected by possible 
uncertainties in the absolute photometric zero-point. These advantages 
become even more compelling when dealing with large cluster samples, and indeed,  
$\omega$ Cen RRL provides the largest cluster sample after M3 and M62.     
The data in Fig.~\ref{fig:omegacenbailey} 
show, from top to bottom the amplitudes in {\it B\/}, {\it V\/} 
and {\it I\/} band. The two solid lines overplotted on the RRab variables display the 
analytical relations for OoI and OoII clusters derived by \citet{cacc2005}, 
while the solid line plotted over the RRc variables is the analytical relation 
for OoII clusters derived by \citep{kunder2013c}.

The majority of the RRab of $\omega$ Cen lie along the OoII 
locus for periods longer than $\sim$0.6 days, and along the OoI locus 
for shorter periods. On the other hand, RRab with periods longer than 
0.80 days show, at fixed period, amplitudes that are systematically 
larger than typical for OoII clusters. Moreover, they also display a 
long-period tail not present in typical OoII clusters.
The same distribution has already been observed in the $V$-band Bailey 
diagram provided by  \citet{clementrowe2000,kaluzny_etal2004}.
More interestingly, there is evidence that a significant 
fraction (79\%) of candidate Blazhko RRLs (22 out of 28) have periods 
shorter than 0.6 days. This finding further supports the evidence 
originally brought forward by \citet{smith1981} concerning the lack of 
Blazhko RRLs with a period longer than $\approx$0.7 days.  
Note that the Blazhkocity \citep{kunder2013c} among the RRab of 
$\omega$ Cen with periods shorter than 0.6 days is of the order 
of 46\%, thus suggesting that $\omega$ Cen is a cluster with a
Blazhkocity that is on average 50\% larger than other GGCs.  
However, this finding could be the consequence that time series data 
of GGCs do not cover with the appropriate cadence large time intervals 
\citep{jurcsik2012}.

The above findings together with similar empirical evidence concerning the 
precise position of RRd variables \citep{coppola2015} sheds 
new light on the topology of the RRL instability strip, and in particular, 
on the color/effective temperature range covered by the different kind of 
pulsators.         

The RRc (light blue squares) plotted in Fig.~\ref{fig:omegacenbailey} display the typical 
either "hairpin" or "bell" shape distribution. 
The OoII sequence from \citet{kunder2013c} appears to be, at fixed pulsation period, 
the upper envelope of the RRc distribution. Moreover, they seem to belong to two 
different sub-groups (if we exclude a few long- and short-period outliers): 
a) short-period---with periods ranging from $\sim$0.30 to $\sim$0.36 days 
and visual amplitudes ranging from a few hundredths of a magnitude to a few tenths;
b) long-period---with periods ranging from $\sim$0.36 to $\sim$0.45 days
and amplitudes clustering around $AV\sim$0.5 mag. 
With the only exception of the metal-rich clusters \ngc{6388} 
\citep{pritzl2002} and \ngc{6441} \citep{pritzl2001}, and
with V70 in M3 \citep{jurcsik2012}, $\omega$ Cen is the only GGC 
where long-period RRc are found \citep{catelan2004b}.
Theoretical and empirical evidence indicates that the RRc period distribution 
is affected by metallicity \citep{dallora03}. An increase in metal 
content causes a steady decrease in the pulsation period \citep{bono1997}.  
The above evidence seems to suggest that the dichotomous distribution of 
RRc might be the consequence of a clumpy distribution in metal abundance 
(see Section~\ref{chapt_metallicity_omega}).
The reader interested in detailed insights on the metallicity dependence of 
the RRLs position in the Bailey diagram is referred to \citet{navarrete_etal2015}.

\begin{figure*}[t]
\centering
\figurenum{7}
\includegraphics[width=12cm]{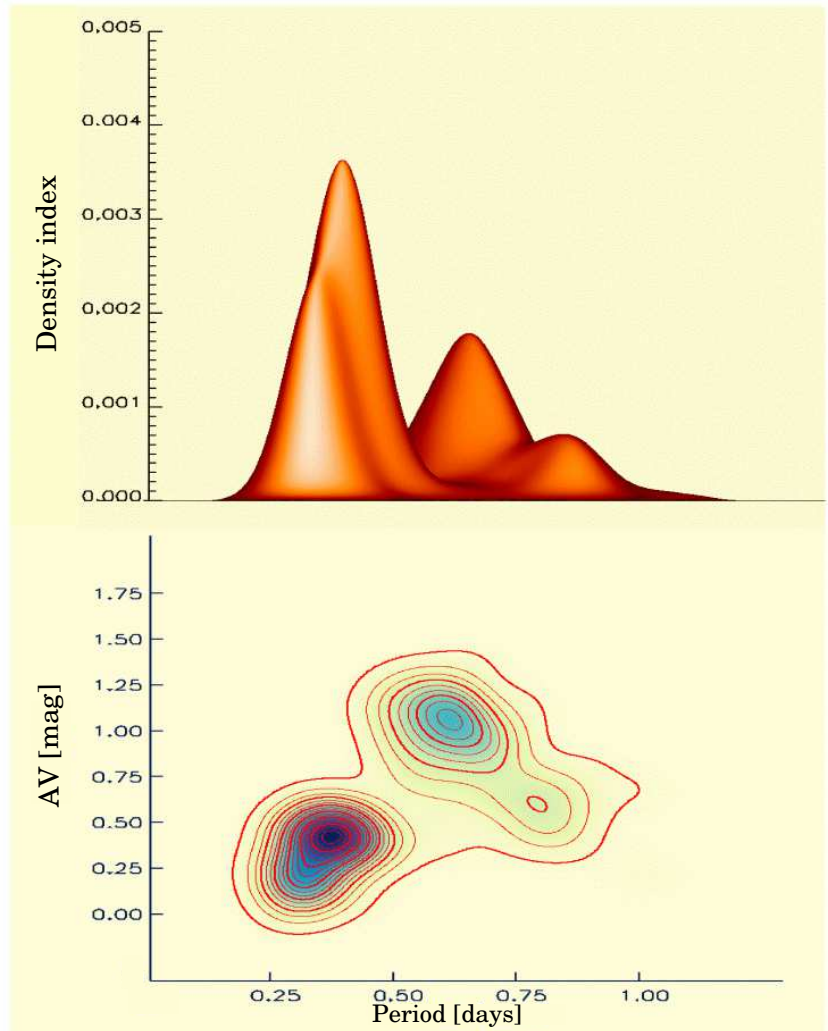}
\vspace*{1.0truecm}
\caption{
Top: 3D Bailey diagram for candidate $\omega$ Cen RRLs: 
pulsation period in days, {\it V\/}-band amplitude and the Z-axis in 
arbitrary units. 
Bottom: Same as the top, but the view is from the top.}
\label{fig:baley3d_omega}
\end{figure*}

To further constrain the fine structure of the Bailey diagram we plotted the 
same variables in a 3D plot. The distribution was smoothed with a Gaussian 
kernel. The top panel of Fig.~\ref{fig:baley3d_omega} shows that the distribution is far from being 
homogeneous, and indeed, both the RRc and the RRab variables show double 
secondary peaks in the shorter and in the longer period range, respectively. 
This evidence is further supported by the iso-contours plotted in the bottom 
panel of the same figure. The iso-contours were estimated running a Gaussian 
kernel, with unit weight, over the entire sample. In this panel the 
long-period of RRab variables can also be easily identified. 

Despite the fact that the Bailey diagram for $\omega$ Cen RRL shows some peculiarities,  
these results fully support the OoII classification for $\omega$ Cen 
RRL suggested by \citet{clementrowe2000} and more recently by 
\citet{navarrete_etal2015}.

\subsection{Luminosity amplitude ratio}\label{chapt_amplratio}

The amplitude ratios are fundamental parameters together with the periods 
and the epoch of a reference phase (luminosity maximum, mean magnitude) for 
estimating the mean magnitude of variable stars using template light curves. 
This approach provides mean magnitudes with a precision of a few hundredths 
of a magnitude from just a few phase points \citep{jones1996,soszynski2005,inno2015}.
Two key issues that need to be addressed in using the amplitude ratios are 
possible differences between RRab and RRc variables and the metallicity dependence 
\citep{inno2015}.  The $\omega$ Cen RRLs play a key role in this context, 
for both the sample size and the well known spread in iron abundance.

\begin{figure*}[t]
\centering
\figurenum{8}
\includegraphics[width=12cm]{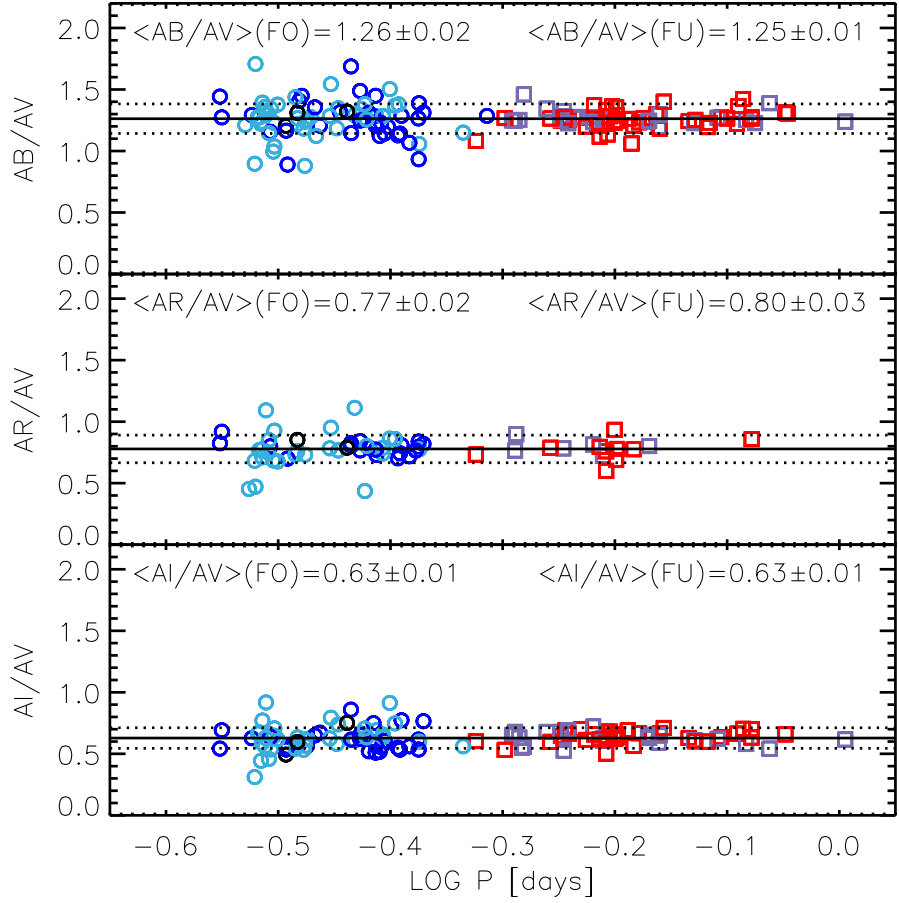}
\vspace*{1.5truecm}
\caption{Top: Ratio between the {\it B\/}- and the {\it V\/}-band amplitude versus 
the pulsation period for candidate $\omega$ Cen RRLs. Metal-poor 
and metal-rich variables (threshold at [Fe/H] = --1.7) are shown with
different colors. Light blue and red symbols display metal-poor RRc 
and RRab, while blue and violet symbols mark metal-rich RRc and RRab, respectively. 
Black symbols are for variables with no [Fe/H] estimate.
The iron abundances come from R00 and S06 and the adopted values 
are listed in column~4 of Table~\ref{tbl:omegametallicity} 
(see text for more details concerning iron abundances).
The average value of the amplitude ratio for RRab and 
RRc and the error on the mean are also labeled. The solid line shows the 
mean value of the global sample (All), while the dashed lines display 
the standard deviation.
Middle: Same as the top, but for the amplitude ratio between {\it R\/}- and {\it V\/}-band.
Bottom: Same as the top, but for the amplitude ratio between {\it I\/}- and {\it V\/}-band.
}
\label{fig:amplratio}
\end{figure*}

Following the same approach adopted by \citet{kunder2013c} and 
\citet{stetson2014}, we estimated the amplitude ratios in different 
bands. Fig.~\ref{fig:amplratio} shows the mean values of the amplitude 
ratios: $AB$/$AV$ (top), $AR$/$AV$ (middle) and 
$AI$/$AV$ (bottom) of $\omega$ Cen RRLs. We included only variables 
with the best-sampled light curves.  We have quantified the goodness of the 
sampling of the light curve with a quality parameter, based on
the number of phase points, the presence of phase gaps and the uncertainties
in the magnitudes of the individual phase points. The paucity of RRab variables in the middle panel 
is due to the fact that our {\it R\/}-band photometry was mostly collected during 
two single nights. Therefore, the {\it R\/}-band light curves of long-period RRLs 
are not well-sampled. 

The amplitude ratios were estimated using the bi-weight to remove the outliers 
\citep{beers1990,fabrizio2011,braga2015}. 
The individual values for RRab, RRc and for the global (All) samples are 
listed in Table~\ref{tab:amplratio} together with their errors and standard 
deviations. The errors account for the uncertainty in the photometry and in 
the estimate of luminosity maxima and minima. 
Estimates listed in Table~\ref{tab:amplratio} and plotted in Fig.~\ref{fig:amplratio} indicate 
that there is no difference, within the errors, between the RRab and RRc amplitude ratios. 
Moreover, the data in Fig.~\ref{fig:amplratio} show no clear dependence on the 
metal content: indeed metal-rich ([Fe/H] > --1.70, blue and violet symbols) 
and metal-poor ([Fe/H] $\leq$ --1.70, light blue and red symbols) display quite similar amplitude ratios.  

In passing we note that the RRc amplitude ratios have standard deviations that 
are larger than the RRab ones. The difference is mainly caused by the fact that 
short-period RRc are characterized by low-amplitudes and small amplitude changes 
cause larger fractional variations. The standard deviations of 
RRab and RRc attain almost identical values if we consider only variables 
with {\it V\/}-band amplitudes larger than 0.35 mag. The difference is mainly caused 
by small uncertainties in the luminosity variation amplitudes causing a larger spread in the 
amplitude ratios.

In summary, the amplitude ratios of $\omega$ Cen RRLs agree quite 
well with similar estimates for other GGCs available in the literature 
\citep{dicriscienzo2011,kunder2013c,stetson2014}. To further characterize the possible dependence on 
metal content of the amplitude ratios we also estimated $AV$/$AI$, 
$AB$/$AI$ and $AB$/$AR$. The means, their errors and standard 
deviations are also given in Table~\ref{tab:amplratio}. We found that 
the current ratio $AV$/$AI = 1.60 \pm 0.02$ agrees quite well with the 
estimate provided by \citet[][see their Tables~3 and 4]{kunder2013c}. There is 
one outlier \ngc{3201}, but this cluster contains only four RRc. The ratio  
$AB$/$AI = 2.00 \pm 0.02$ is also in reasonable agreement with 
literature values. There are two outliers, namely \ngc{6715} and \ngc{3201}, 
that are classified as Oo Int clusters (see Section~\ref{chapt_comparison_oosterhoff}). 
The $AB$/$AV\sim1.25$ ratio agrees well with literature values, but slightly larger 
values have been found for M22 and NGC~4147 ($AB$/$AV\sim1.37$). 
Finally, the ratio $AB$/$AR$ of the RRL in $\omega$ Cen is, within the errors, 
the same as in M4 \citep{stetson2014}.

On the whole the above findings indicate that F and FO amplitude ratios do not 
depend on the metal content in the range covered by RRL in $\omega$ Cen 
([Fe/H]=--2.4$\div$--1.0) and in the other clusters considered 
([Fe/H]=--2.4$\div$--1.2, \citealp{harris1996,kunder2013c}).

\section{The RR Lyrae in the Color-Magnitude Diagram}\label{chapt_position_omega}

\begin{figure*}[t]
\centering
\figurenum{9}
\includegraphics[width=11cm]{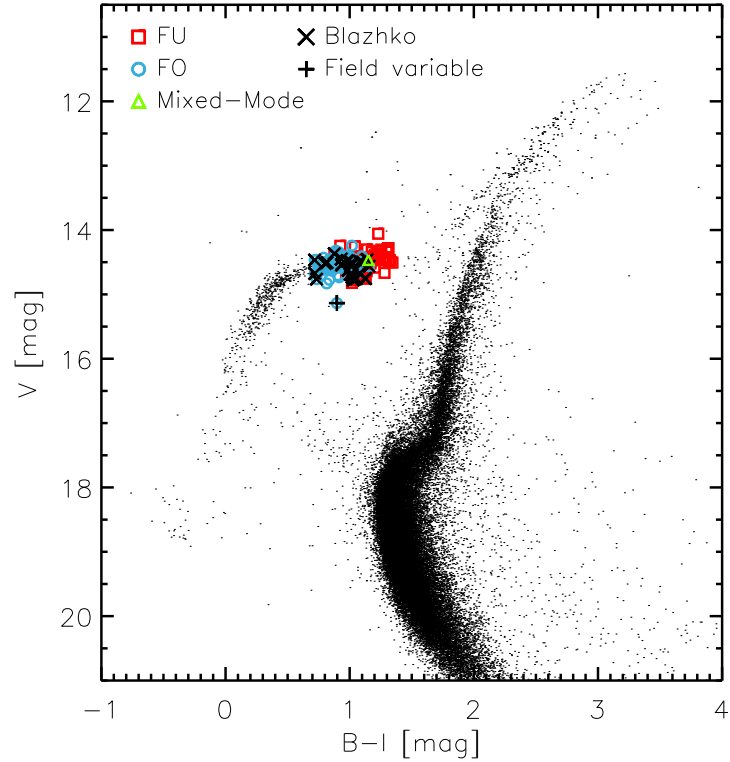}
\vspace*{1.0truecm}
\caption{Optical ({\it V\/},\bmi) color-magnitude diagram of $\omega$ Cen. 
Light blue circles and red squares mark FO and F RRLs. The candidate RRd variable V142
is marked in green. The candidate FO field variable (V168) is marked with a black 
plus, while the black crosses display candidate Blazhko stars.}
\label{fig:omegacen_cmd_bvi}
\end{figure*}

The current photometry allowed us to derive an accurate CMD 
covering  not only the bright region typical of RGB and AGB stars 
(V$\sim$11-12 mag), but also $\sim$3 magnitudes fainter than the main 
sequence turn-off region. Fig.~\ref{fig:omegacen_cmd_bvi} shows the 
optical {\it V\/}, \bmi\ CMD of $\omega$ Cen.
The stars plotted in the above CMD have been selected using the
photometric error ($\sigma_V\sim$0.03, $\sigma_{B-I}\sim$0.04 mag),
the $\chi$ parameter (< 1.8), quantifying the deviation between 
the star profile and the adopted Point Spread Function (PSF), and the sharpness ($\bigl| sha \bigr|$ $<$ 0.7)
quantifying the difference in broadness of the individual stars compared 
with the PSF. In passing we note that PSF photometry of 
individual images is mandatory to improve the precision of individual 
measurements of variable stars. The identification and fitting of faint sources 
located near the variable stars provides an 
optimal subtraction of light contamination from neighboring 
stars.

On top of the cluster photometry, Fig.~\ref{fig:omegacen_cmd_bvi} also 
shows the 170 out of the 195 RRLs for which we estimated both 
{\it B\/}-, {\it V\/}- and {\it I\/}-band mean magnitudes. 
The light blue, red and green symbols display RRc, RRab and 
the candidate RRd variable. The RRc are located, as expected, on 
the blue (hot) side of the instability strip, while the RRab are in the 
red (cool) region of the instability strip \citep{bono1997c}. The crosses 
mark candidate Blazhko variables. 
The black plus sign identifies a candidate RRc field variable---V168---with a mean 
visual magnitude that is $\sim$0.6 mag fainter than cluster variables. 

\begin{figure*}[t]
\centering
\figurenum{10}
\includegraphics[width=15cm]{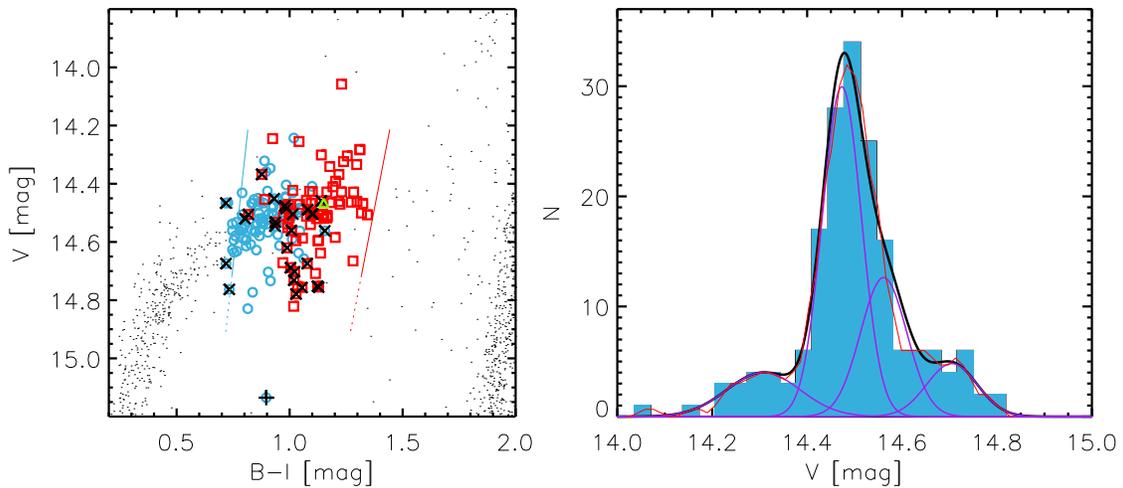}
\vspace*{1.0truecm}
\caption{Left: Same as Fig.~\ref{fig:omegacen_cmd_bvi}, but zoomed on the RRL instability strip. 
The blue line shows the predicted First Overtone Blue Edge (FOBE), while the red one the predicted 
Fundamental Red Edge (FRE) of the instability strip \citep{marconi2015}. Right:  
distribution of the {\it V\/} magnitudes of RRLs. The red curve shows the smoothed histogram. The four 
components of the multi-Gaussian fit are plotted in purple, while the black curve is the sum of 
the multi-Gaussian fit.}
\label{fig:omegacen_cmd_bvi_rrl}
\end{figure*}

To further define the range in magnitude and colors covered by 
cluster RRLs, the left panel of Fig.\ref{fig:omegacen_cmd_bvi_rrl} shows a 
zoom across the instability strip. The blue and the red lines display the 
predicted hot (blue) edge for FO pulsators (FOBE) and the cool (red) edge 
for F pulsators (FRE). Note that the predicted edges are based on the 
analytical relations provided by \citet{marconi2015} (see their Table~5). 
We assumed a metal content $\log{Z}$ = 0.0006 and an $\alpha$-enhanced 
chemical mixture ([$\alpha$/Fe]=0.4). This means an iron abundance of 
[Fe/H]=--1.81\footnote{The reader is referred to \citet{pietrinferni2006}
and to the BASTI data base (\url{http://albione.oa-teramo.inaf.it})
for a more detailed discussion concerning the evolutionary framework 
adopted in constructing the pulsation models.}  
These iron and $\alpha$-element abundances are consistent with the 
peak in the metallicity distribution of evolved stars in $\omega$ Cen 
based on recent spectrophotometric \citep{calamida_etal2009} and 
spectroscopic \citep{johnson_e_pilachowski2010} measurements.
The agreement between theory and observations is remarkable if 
we take account for the theoretical  
and empirical uncertainties at the HB luminosity level.
The former include a $\sim$50 K uncertainty on the temperature
of the computed models, 
taking account for the adopted step in temperature \citep{dicriscienzo2004}, 
plus uncertainties in 
color-temperature transformations ($\sigma_{B-I}\approx$0.05 mag). 
The empirical uncertainty on both the FRE and the FOBE is 
$\sigma_{B-I}\sim$0.05 mag.
Note that the possible occurrence of differential reddening 
($\Delta$E(\bmv)=0.04 mag, \citep{monibidin2012}) mainly 
causes an increase in the photometric dispersion across the 
boundaries of the instability strip.  
The distribution of the RRLs inside the instability strip shows 
two interesting empirical features worth being discussed in more 
detail.

{\em Magnitude distribution}---To provide firm estimates of spread
in visual magnitude  of the $\omega$ Cen RRLs we performed an analytical
fit of the observed distribution. The right panel of
Fig.~\ref{fig:omegacen_cmd_bvi_rrl} shows the observed {\it V\/} magnitude
distribution as a blue histogram. To overcome deceptive uncertainties
in the criteria adopted to bin the data, we smoothed the distribution
assigning to each RRL a Gaussian kernel \citep{dicecco2010} with a
$\sigma$ equal to the intrinsic error of the mean magnitude. The red
curve was computed by summing the individual Gaussians over the entire
dataset. The main peak appears well defined and located at
$\sim$14.5 mag.
To provide a more quantitative analysis, we fit the smoothed
magnitude distribution with four Gaussian functions (purple curves). Note that 
the number of Gaussians is arbitrary: they were included only to 
minimize the residuals between analytical and observed distribution. 
The black solid curve shows the sum of the four Gaussians 
over the entire 
magnitude range. The data listed in Table~\ref{tab:gaussfit} indicate that 
the two main peaks are located at V$\sim$14.47 and V$\sim$14.56 mag and 
include a significant fraction of the entire RRL sample, $\approx$51\% and 
$\approx$25\%, respectively. The fainter and the brighter
peaks are located at V$\sim$14.71  and V$\sim$14.31 mag and roughly
include $\approx$11\% and $\approx$13\% of the RRL entire
sample. This suggests the metal-rich and the metal-poor tail
produce only a minor fraction of RRLs. The above spread in optical
magnitude indicates, for a canonical $M_V^{RR}$~vs~[Fe/H]
relation \citep{bono1997a}, that $\omega$ Cen RRLs cover a range in
iron abundance of the order of 1.5 dex (see also \S~\ref{chapt_metallicity_omega}).
 
{\em Blazhko RR Lyrae}---The data in the left panel 
of Fig.~\ref{fig:omegacen_cmd_bvi_rrl}, indicate 
that a significant fraction (39\%) of candidate Blazhko RRLs belongs to the 
fainter peak (V$\geq$14.6 mag). 
Preliminary evidence of clustering in magnitude and in color of Blazhko RRLs, 
has been found in M3 by \citet{catelan2004a}, but a Kolmogorov-Smirnov test gave
negative results. We have performed the same test on the \bmv and \bmi color
distributions of Blazhko RRLs versus RRab and RRc variables. We found that the 
probability the color distribution of Blazhko RRLs being equal to the 
color distribution of RRab and RRc is on average smaller than 1\%.   
Moreover, we also confirm the preliminary empirical evidence based 
on the Bailey diagram (see \S~\ref{chapt_amplratio}): 
they are mainly located between the FO and the F instability region. 
The above finding suggests that candidate Blazhko RRLs in $\omega$ Cen 
attain intermediate colors/temperatures. 

Moreover, the difference in mean visual magnitude between the fainter 
(V$\geq$14.6 mag) and the brighter (V$<$14.6 mag) sample is also suggesting 
that the former ones are slightly more metal-rich. 
This working hypothesis is supported by metallicity 
estimates based on spectrophotometric indices \citep{rey2000} suggesting, 
for fainter and brighter Blazhko RRLs, mean metallicities of --1.4$\pm$0.3 
and --1.8$\pm$0.1 dex (see, e.g., \S~\ref{chapt_metallicity_omega}). Metallicity 
estimates based on spectroscopic measurements \citep{sollima2006} support the 
same finding, and indeed the mean iron abundances for fainter and brighter 
Blazhko RRLs  are --1.2$\pm$0.1 and --1.7$\pm$0.2 dex, respectively. 
In passing, we also note that empirical evidence indicates that 
the Blazhko phenomenon occurs with higher frequency in more metal-poor 
environments \citep{kunder2013c}. Homogeneous and accurate spectroscopic 
iron abundances are required to further investigate this interesting 
preliminary result.

\section{Comparison with RR Lyrae in globulars and in dwarf galaxies}\label{chapt_comparison_oosterhoff}

The large number of RRLs in $\omega$ Cen allows us to perform a detailed 
comparison with pulsation and evolutionary properties of RRLs in nearby 
stellar systems. To overcome thorny problems caused by small number statistics 
we selected, following \citet{fiorentino2015}, only GGCs hosting at least 
three dozen (35) RRLs. They are 16 out of the $\sim$100 GGCs hosting RRLs 
\citep{clement2001}. 
To characterize the role played by the metallicity in shaping their pulsation 
properties they were divided, according to their metal content \citep{harris1996}, 
into four different groups: 

OoI\footnote{Oosterhoff type I clusters: \ngc{5272}, \ngc{5904}, \ngc{6121}, \ngc{6229},  
\ngc{6266}, \ngc{6362}, \ngc{6981}.}---including 402~RRab, 6~RRd 
and 165~RRc with iron abundances ranging from [Fe/H]=--1.00 to --1.50;   

OoInt\footnote{Oosterhoff intermediate clusters: IC~4499, \ngc{3201}, 
\ngc{6715}, \ngc{6934}, \ngc{7006}.}---including 324~RRab, and 50~RRc with iron abundances ranging from 
[Fe/H]=--1.50 to --1.65;   

OoII\footnote{Oosterhoff type II clusters: \ngc{4590}, \ngc{5024}, \ngc{5286}, \ngc{7078}.}---including 
111~RRab, 28~RRd and 111~RRc with iron abundances ranging 
from [Fe/H]=--1.65 to --2.40.   

OoIII\footnote{Oosterhoff type III/Oosterhoff type 0 clusters: \ngc{6388}, \ngc{6441}.}---including 
60~RRab, 1~RRd and 41~RRc belonging to the two 
metal-rich globulars \citep{pritzl2001,pritzl2002,pritzl2003} 
\ngc{6388} ([Fe/H]=--0.55) and \ngc{6441} ([Fe/H]=--0.46). 
Note that we did not include the RRLs recently identified in Terzan 10 
and in 2MASS-GC 02 by \citet{alonsogarcia2015}, since these two clusters 
still lack accurate spectroscopic measurements of the iron abundance.  

The data for RRLs in GGCs were complemented with similar data for
RRLs in nearby gas-poor stellar systems, namely dwarf spheroidal (dSph)
and Ultra Faint Dwarf (UFD) galaxies. Note that we did not apply any 
selection criterion on the number of RRLs in building up this sample. 
We ended up with a sample of 1306~RRab, 50~RRd and 369~RRc with iron 
abundances ranging from [Fe/H]=--2.6 to [Fe/H]=--1.4 
\citep{kirby2013,mcconnachie2012,fabrizio2015}.

The double-mode variables---RRd---pulsate simultaneously in two 
different radial modes, typically F and FO. 
However, the latter is, with only a few exceptions (V44 in M3, 
\citealp{jurcsik2015}), the main mode. 
However, they were not plotted in the Bailey diagram, since
the separation of F and FO light curves does require very accurate and
well sampled light curves \citep{coppola2015}. They were also excluded
from the period distribution, but included in the RRL population ratio,
i.e., the ratio between the number of RRc and the 
total number of RRLs ($N_c/N_{tot}$). 
We plan to provide a more detailed analysis of RRd variables 
in a follow-up paper.

In this context it is worth mentioning that the RRLs that in the 
Clement catalog are classified as second overtones---RRe---were 
treated as RRc variables. Theoretical and empirical evidence indicates 
that the steady decrease in the pulsation period of RRc variables is 
mainly caused by a steady increase in metal content \citep{bono1997}. 
Note that the conclusions concerning the comparison between RRLs 
in $\omega$ Cen and in the other stellar systems are minimally 
affected by the inclusion of double-mode and possible candidate 
second overtone RRLs.  

\begin{figure*}[t]
\centering
\figurenum{11}
\includegraphics[width=12cm]{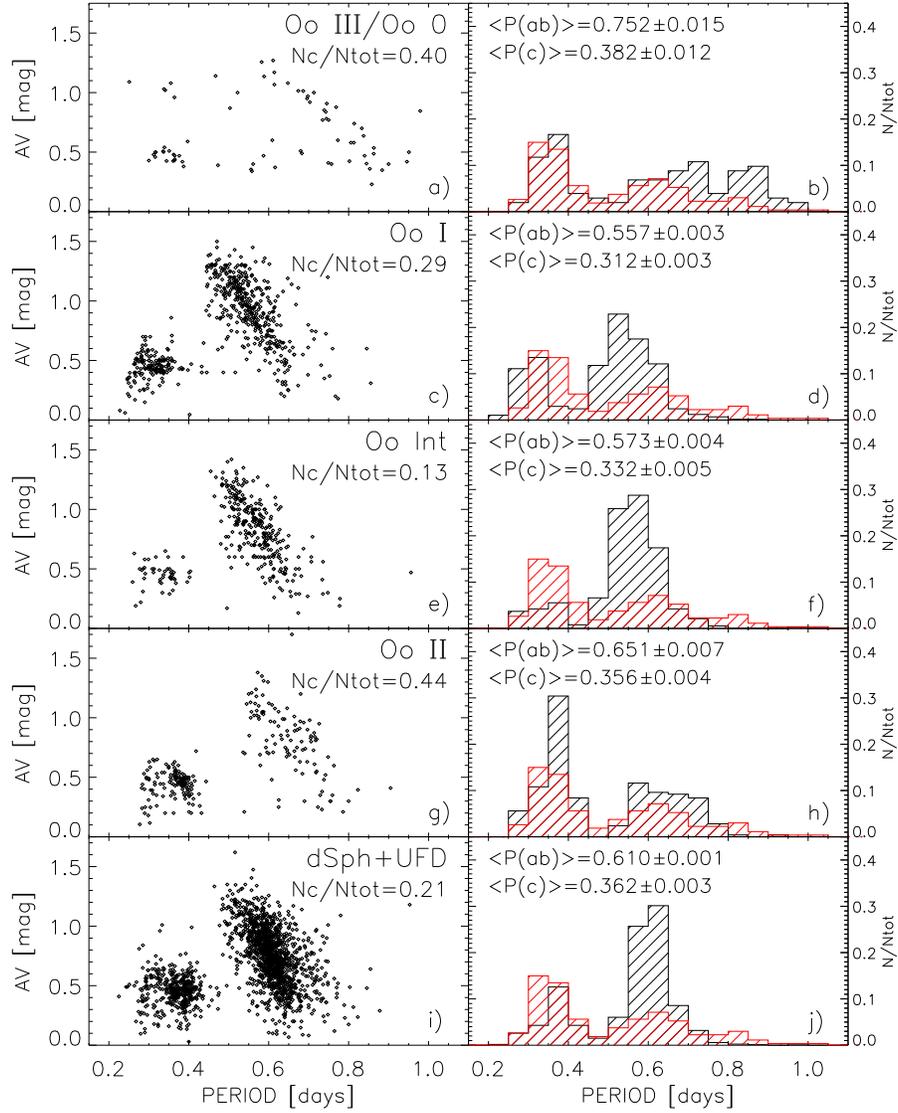}
\vspace*{1.0truecm}
\caption{
Panel a): Bailey diagram of the RRLs in the two OoIII/Oo0 
metal-rich GGCs \ngc{6388} and \ngc{6441}. Panel b): Comparison between 
the period distribution of RRLs plotted in Panel a) (black shaded area) 
and in $\omega$ Cen (red shaded area). The mean periods of RRab and RRc 
and the population ratio (number of RRc over the total number of RRLs) 
of the two GCs are also labeled.
Panel c) and d): Same as a) and b), but for OoI GGCs with more than 35 RRLs.
Panel e) and f): Same as a) and b), but for OoInt GGCs with more than 35 RRLs.
Panel g) and h): Same as a) and b), but for OoII GGCs with more than 35 RRLs.
Panel i) and j): Same as a) and b), but for dwarf spheroidals (dSph)
and Ultra-Faint Dwarf (UFD) galaxies.}
\label{fig:histoper_bailey_comparison2}
\end{figure*}

We estimated the diagnostics adopted to describe the Oosterhoff dichotomy:
mean RRab and RRc periods and RRL population ratio for the stellar 
systems considered here, and their values are listed in Table~\ref{tab:oosterhoff_comparison}
together with their uncertainties. We have already mentioned in Section~\ref{par:bailey} 
that $\omega$ Cen RRLs follow quite closely OoII clusters. However, data listed in 
this table together with the amplitudes and the period distributions plotted 
in Fig.~\ref{fig:histoper_bailey_comparison2} display 
several interesting trends worth being discussed. 

{\em i)}---Linearity---The mean periods display a steady increase when moving 
from more metal-rich to more metal-poor stellar systems. The exception in 
this trend is given by the two metal-rich Bulge clusters (\ngc{6388}, \ngc{6441}).  
They are at least a half dex more metal-rich than OoI clusters, but their mean 
periods are from $\sim$25\% (RRc) to $\sim$35\% (RRab) longer. In passing we 
note that the above findings suggest that metal-rich globulars hosting RRLs 
belong to the Oosterhoff type 0 clusters instead of the OoIII group.
This is the reason why their amplitudes and periods were plotted on top of 
Fig.~\ref{fig:histoper_bailey_comparison2} (see also the discussion in 
Section~\ref{chapt_final}).   
On the other hand, the RRL population ratio shows a nonlinear trend, and 
indeed the OoInt clusters display a well defined minimum when compared 
with OoI, OoII and OoIII/Oo0 clusters. The decrease ranges from more than a 
factor of two with OoI to more than a factor of three with OoII and OoIII/Oo0 
clusters. The RRLs in dwarf galaxies appear to attain values typical of stellar systems 
located between OoInt and OoII clusters. Note that the RRc mean period attains 
very similar values in dwarfs, in OoII clusters and in $\omega$ Cen, thus 
suggesting a limited sensitivity of this parameter in the more metal-poor 
regime.

{\em ii)}---Nature---The results mentioned in the above paragraph open the 
path to a long-standing question concerning the nature of $\omega$ Cen, 
i.e., whether it is a massive globular cluster or the former core of a 
dwarf galaxy. To further investigate this interesting issue we performed  
a more quantitative comparison between RRLs in $\omega$ Cen and in the 
aforementioned gas-poor stellar systems.  The data in the left panels of 
Fig.~\ref{fig:histoper_bailey_comparison2} display two clear features: 
a)  $\omega$ Cen and dwarf galaxies lack of High Amplitude Short Period (HASP) 
RRLs, i.e., F variables with P$\lesssim$0.48 days and $AV$ $>$ 0.75 mag  
\citep{stetson2014b}. 
Empirical and theoretical evidence indicates that they become more and more 
popular in stellar systems more metal-rich than [Fe/H]$\approx$ --1.4/--1.5 
\citep{fiorentino2015}.
Therefore, the paucity of HASPs in $\omega$ Cen is consistent with 
previous metallicity estimates available in the literature 
\citep{calamida_etal2009,johnson_e_pilachowski2010}, and with the current 
metallicity estimates (see Section~\ref{chapt_metallicity_omega}).
We estimated the marginals of the Bailey diagrams plotted in the left panels 
of the above figure plus $\omega$ Cen and the $\chi^2$ analysis indicates 
that the latter agrees with OoII clusters at the 94\% confidence level. 
The agreement with the other Bailey diagrams is either significantly 
smaller (39\%, OoIII/Oo0; 42\%, dwarfs) or vanishing (OoI, OoInt).

On the other hand, the comparison of the period distributions plotted in the 
right panels of Fig.~\ref{fig:histoper_bailey_comparison2} clearly 
display that $\omega$ Cen is similar to an OoII cluster.  Moreover, 
RRLs in $\omega$ Cen and in dwarf galaxies also display similar 
metallicity distributions. However, the coverage of the RRL instability 
strip in the former system appears to be more skewed toward the FO region 
than toward the F region as in the latter ones.  The above difference is 
further supported by the stark difference in the population ratio and in 
the peaks of RRab and RRc period distributions. 
We also performed the $\chi^2$ analysis of the period distributions plotted 
in the right panels of Fig.~\ref{fig:histoper_bailey_comparison2} and we found 
that RRLs in $\omega$ Cen agree with OoII clusters at the 80\% confidence level. 
The agreement with the other samples is either at a few percent level or 
vanishing (dwarfs). Therefore, the working 
hypothesis that $\omega$ Cen is the core remnant of a spoiled dwarf 
galaxy \citep{zinnecker1988,freeman1993,bekkifreeman2003} does not find 
solid confirmation by the above findings. 
This result is somehow supported by the lack of firm signatures of tidal 
tails recently found by \citep{fernandeztrincado2015,fernandeztrincado2015b} 
using wide field optical photometry covering more than 50 $deg^2$ around 
the cluster center.

{\em iii)}---Nurture---$\omega$ Cen RRLs display a well defined long-period 
tail (P$>$0.8 days) that is barely present in the RRL samples of the other systems. The
exception is, once again, given by the two metal-rich globulars hosting 
RRLs, namely \ngc{6388} and \ngc{6441}. A detailed analysis 
of the HB luminosity function is beyond the aim of the current investigation, 
however, we note that $\omega$ Cen and the two Bulge clusters share an 
indisputable common feature, i.e., the presence in the HB luminosity function 
of a well extended blue tail. 
This suggests us that its presence is more nurture than nature. 
The environment, and in particular, the high central density, might play a 
crucial role in the appearance of the blue tail, and in turn in the appearance 
of long-period RRLs \citep{castellani2006}. Indeed, it has been suggested 
\citep{castellani_etal2007,latour2014} that either binarity or stellar encounters might 
explain the presence of extended blue tails, and in turn, an increased fraction 
of blue HB stars evolving from the blue to the red region of the CMD.
However, it is worth noting that the 
above evidence is far from taking account of the current empirical evidence, 
and indeed the metal-intermediate \citep[{[}Fe/H{]}=--1.14][]{carretta2009b}
globular \ngc{2808} hosts 11 RRab variables, but they have periods shorter than 
0.62 days \citep{kunder2013a}.
It has also been suggested that a possible spread in helium abundance might 
also take account for the HB morphology in $\omega$ Cen \citet{tailo2016}, 
and in turn of the period distribution of RRLs. However, the increase in 
helium content is degenerate with possible evolutionary effects 
\citep{marconi2011} and we still lack firm conclusions.
The reader interested in a recent detailed discussion concerning the Oosterhoff 
dichotomy and the HB morphology is referred to \citet{janglee2015}.  
 
Finally, we would like to underline that the above results strongly support
the idea that only a limited number of GCs are good laboratories to understand 
the origin of the Oosterhoff dichotomy. The main limitations being statistics 
and environmental effects. This evidence further suggests that 
the metallicity is the main culprit in shaping the above empirical evidence, 
while the HB luminosity function appears to be the next more plausible candidate.
In passing, we also mention that a steady increase in helium content has also 
been suggested to take account of the extended blue tail in Galactic globulars 
\citep[\ngc{2808},][]{dantona2005}. The increase in helium content 
causes a steady increase in the pulsation period of both RRc and RRab variables 
\citep{marconi2011}. Firm constraints require detailed sets of synthetic 
HB models accounting for both the HB morphology and the period distribution 
\citep{salaris2013,sollima2014,savino2015}. 
We plan to investigate this issue in a forthcoming paper, 
since $\omega$ Cen is the perfect laboratory to constrain the transition from 
RRLs to TIICs.

\section{RR Lyrae diagnostics}\label{chapt_rrl_diag}
\subsection{Period-Luminosity relations}\label{chapt_empirical_pl_omega}

\begin{figure*}[t]
\centering
\figurenum{12}
\includegraphics[width=12cm]{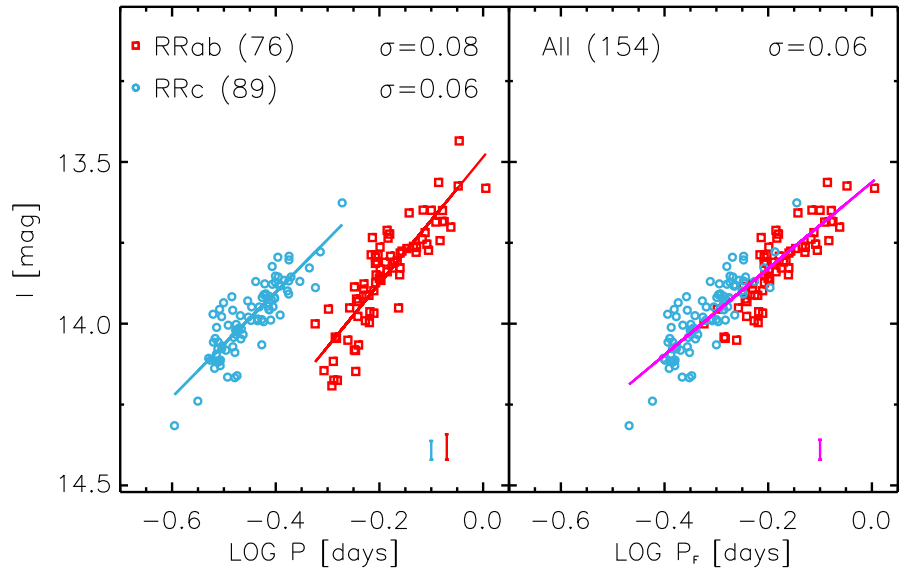}
\caption{Left: Empirical {\it I\/}-band Period-Luminosity (PL) relation for 
$\omega$ Cen RRLs. Light blue and red squares mark RRc and RRab variables.  
The light blue and the red lines display the linear fits, while the vertical 
bars show the standard deviations, $\sigma$, of the fits. 
The number of variables adopted in the fits are also labeled.
Right: Same as the left, but for the global (RRc$+$RRab) RRL sample. 
The periods of RRc variables were fundamentalized using 
$\log{P_{F}} = \log{P_{FO}} + 0.127$.}
\label{fig:PL_empirical_omega}
\end{figure*}

On the basis of both periods and mean magnitudes measured in 
Section~\ref{par:omegacen_lightcurves} and in Section~\ref{par:perioddistribution},
we estimated the empirical {\it I\/}-band PL relations of $\omega$ Cen RRLs. Following 
\citet{braga2015} and \citet{marconi2015} we evaluated the PL relations for 
RRc, RRab and for the global (All) sample. In the global sample the RRc were 
``fundamentalized'', i.e., we adopted  $\log{P_{F}} = \log{P_{FO}} + 0.127$ 
\citep{ibenhuchra1971,rood1973,cox1983,dicriscienzo2004,coppola2015}. The 
coefficients, their errors and the standard deviations of the empirical PL 
relations are listed in Table~\ref{tbl:omegacen_pl_empirical}. The RRLs 
adopted to estimate the PL relations are plotted in 
Fig.~\ref{fig:PL_empirical_omega}.

Note that we derived the PL relations only in the {\it I\/}-band because 
theoretical \citep{bono2001,catelan2004,marconi2015} and empirical 
\citep{benko2011,braga2015} evidence indicates that RRLs do not obey to 
a well defined PL relation in the {\it U\/}, {\it B\/} and {\it V\/} band. Moreover,
in the {\it R\/}-band, the dispersion is large ($\sim$0.15 mag) and the slope
is quite shallow ($\sim$-0.5 mag).

The standard deviations plotted in the bottom right corner of 
Fig.~\ref{fig:PL_empirical_omega} and the modest intrinsic error 
on the mean  {\it I\/}-band magnitude discussed in \S~\ref{par:omegacen_lightcurves} 
clearly indicate that the dispersion of the empirical {\it I\/}-band PL relation 
is mainly caused by the spread in metal abundance of $\omega$ Cen RRLs 
(see Section~\ref{chapt_metallicity_omega}). Indeed, pulsation and 
evolutionary predictions \citep{bono2003a,catelan2004,marconi2015} 
indicate that the 
zero-points of the {\it I\/}-band PL relations do depend on metal abundance.
We will take advantage of this dependence to estimate 
individual RRL metal abundances (see \S~\ref{chapt_metallicity_omega}).

\subsection{Period-Wesenheit relations}\label{chapt_empirical_pw_omega}

The Period-Wesenheit (PW) relations, when compared with the PL relations, have the 
key advantage to be reddening-free by construction \citep{vandenbergh1975,madore1982}. 
This difference relies on the assumption that the adopted reddening law is universal 
\citep{bono2010}. The pseudo Wesenheit magnitude is defined as

\begin{equation}\label{eq:wesenheit_definition}
W(X,Y-Z) = X + \dfrac{A_X}{A_Y - A_Z}(Y-Z)
\end{equation}

\noindent
where {\it X\/}, {\it Y\/} and {\it Z\/} are the individual magnitudes and 
$A_X$, $A_Y$ and $A_Z$ are the selective absorption coefficients provided 
by the reddening law \citep{cardelli1989,fitzpatrick2007}.
We have adopted the 
popular reddening law of \citet{cardelli1989} with $R_V=3.06$ 
and $A_B$/$A_{V(Johnson)}$=1.348, $A_V$/$A_{V(Johnson)}$=1.016 $A_I$/$A_{V(Johnson)}$=0.590).
Note that, to match the current optical photometric system \citep{landolt1983,landolt1992},  
the original $R_V$ value ($R_V=3.1$) and the selective absorption ratios provided by 
\citet{cardelli1989} were modified accordingly.

\begin{figure*}[t]
\centering
\figurenum{13}
\includegraphics[width=12cm]{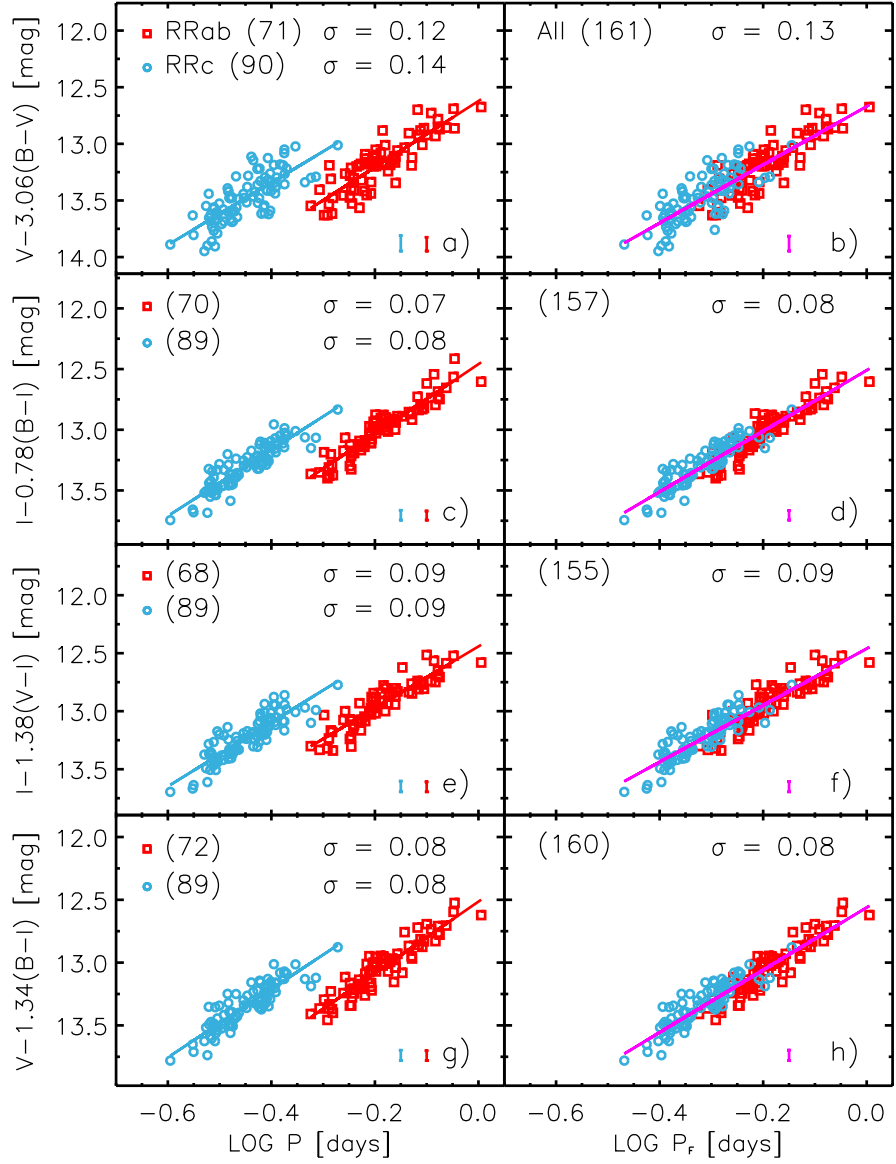}
\caption{
Panel a): Empirical dual-band Period-Wesenheit (PW) ({\it V\/},\bmv) relation for
$\omega$ Cen RRLs. Light blue and red squares mark RRc and RRab variables. 
The light blue and the red lines display the linear fits, while the vertical 
bars show the standard deviations, $\sigma$, of the fits. 
The number of variables adopted in the fits are also labeled.
Panel b): Same as panel a), but for the global (RRc$+$RRab) RRL sample. 
The periods of RRc variables were fundamentalized using 
$\log{P_{F}} = \log{P_{FO}} + 0.127$.
Panels c) and d): Same as a) and b), but for the PW({\it I\/},\bmi) relation. 
Panels e) and f): Same as a) and b), but for the PW({\it I\/},\vmi) relation.
Panels g) and h): Same as a) and b), but for the triple-band PW({\it V\/},\bmi) relation.}
\label{fig:pw_empirical_omega1}
\end{figure*}

Fig.~\ref{fig:pw_empirical_omega1} shows the dual---and the triple---band 
empirical PW relations for $\omega$ Cen RRLs. The coefficients, their 
errors and the standard deviations of the PW relations are listed in 
Table~\ref{tbl:omegacen_pw_empirical}. The slopes of the PW relations 
listed in this Table agree, withing the errors, remarkably well with 
the slope predicted by nonlinear, convective hydrodynamical 
models of RRLs \citep[][see their Tables~7 and 8]{marconi2015}. 
Indeed, the predicted slopes for the metal independent 
PW({\it V\/},\bmv) relations range from --2.8, (FO), to --2.7 (F) and to --2.5 (global), 
while for the metal-dependent PW({\it V\/},\bmi) relations they range from --3.1 (FO),  
to --2.6 (F) and to --2.5 (global). The comparison in the latter case is 
very plausible, since the coefficient of the meatllicity term for the 
PW({\it V\/},\bmi) relations is smaller than 0.1 dex. The predicted slope for 
FO variables is slightly larger, but this might be due to the limited 
sample of adopted FO models. 

The current empirical slopes for the optical PW relations agree quite 
well with similar estimates recently provided by \citet{coppola2015} 
for more than 90 RRLs of the Carina dSph. They found slopes of --2.7 [global, 
PW({\it V\/},\bmv)] and --2.6 [global, PW({\it V\/},\bmi)], respectively. The outcome is 
the same if we take account for the thorough analysis performed by 
\citet{martinezvazquez2015} for the 290 RRLs (clean sample) of Sculptor 
dSph, namely --2.5 [global, PW({\it V\/},\bmi)] and --2.7 [global, PW({\it V\/},\bmi)].   
The reader interested in a detailed discussion concerning the physical 
arguments supporting the universality of the above slope is referred 
to the recent investigation by \citet{lub2016}.

The data in Fig.~\ref{fig:pw_empirical_omega1}
display that the standard deviation of the different PW relations 
steadily decrease if either the effective wavelength of the adopted magnitudes 
increases (see panels c), d) e) and f)) 
and/or the difference in effective wavelength of the 
adopted color increases (see panels d) and h).
Finally, we note that the standard deviations of the PW({\it V\/},$\bmv$) relations 
are systematically larger than the other PW relations. The difference is 
mainly caused by the fact that this PW relation has the largest color 
coefficient (3.06), and in turn, the largest propagation of the intrinsic 
errors on mean colors.

\section{Distance determination}\label{chapt_distance_omega}

The $\omega$ Cen RRLs cover a broad range in metal abundance. This means 
that accurate distance determinations based on diagnostics affected by the 
metal content do require accurate estimates of individual iron abundances 
\citep{delprincipe_etal2006,bono2008a}. The observational 
scenario concerning iron abundances of $\omega$ Cen RRLs is far from being ideal. 
Estimates of the iron abundance for 131 RRLs in $\omega$ Cen were provided by 
\citet[][hereinafter R00]{rey2000} using the {\em hk} photometric index introduced by 
\citet{baird1996}. More recently, \citet[][hereinafter S06]{sollima2006} estimated iron 
abundances for 74 RRLs in $\omega$ Cen using moderately high-resolution spectra 
collected with FLAMES at VLT. These iron abundances are listed in columns 1 and 2 
of Table~\ref{tbl:omegametallicity}. The former sample is in the globular 
cluster metallicity scale provided by \citet[][hereinafter ZW84]{zinnwest1984}. 

\begin{figure*}[t]
\centering
\figurenum{14}
\includegraphics[width=14cm]{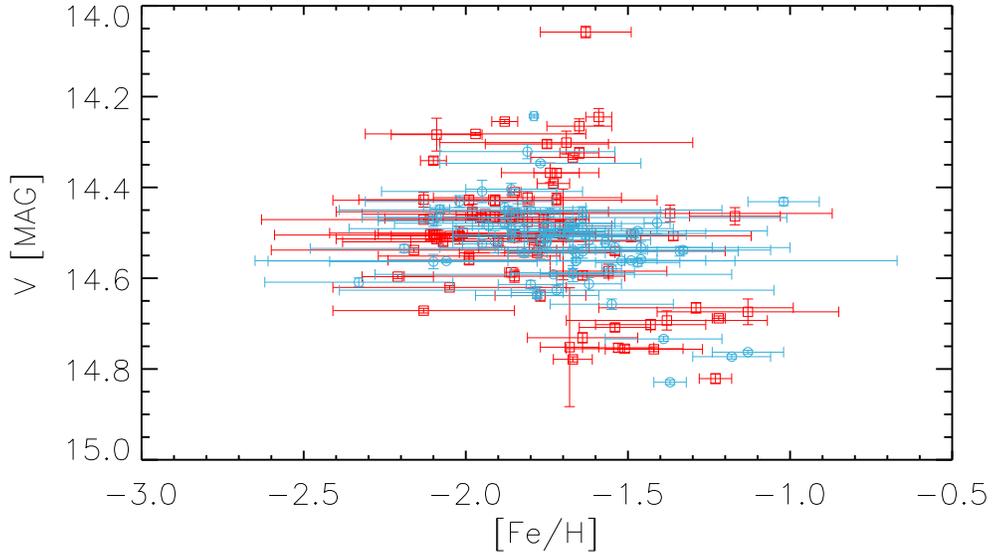}
\vspace*{1.0truecm}
\caption{{\it V\/} vs [Fe/H] distribution of $\omega$ Cen RRLs. We adopted 
iron abundances available in the literature \citep{rey2000,sollima2006}
rescaled into the \citet{carretta2009b} metallicity scale.
The vertical error bars display errors in the mean visual magnitude, 
while the horizontal ones either the intrinsic error (single measurement) 
or the standard deviation (two measurements). See text for more details.}
\label{fig:mv_vs_feh}
\end{figure*}

They were transformed into the homogeneous and accurate metallicity scale 
provided by \citet{carretta2009b} using their linear transformation 
(see their \S~5). The iron abundances provided by S06 were estimated following 
the same approach adopted by \citet{gratton2003,carretta2009b}. 
They were transformed into the Carretta's metallicity scale once accounting 
for the difference in the solar iron abundance ($log{\epsilon_{Fe_{\odot}}}$=7.52 vs 7.54).  
Fortunately enough, the two samples have 52 objects in common. We estimated the 
difference between R00 and S06 and we found 
$\Delta$[Fe/H]=0.18$\pm$0.03 ($\sigma$=0.20). 
We rescaled the R00 to the S06 iron abundances and computed the mean for the 
objects in common (see column 4 in Table~\ref{tbl:omegametallicity}). 
Figure~\ref{fig:mv_vs_feh} shows the entire sample of $\omega$ Cen RRLs 
(153) for which a metallicity estimate is available in the 
[Fe/H]--{\it V\/} plane. A glance at the data  
in this figure shows that the current uncertainties on individual iron abundances 
are too large to provide precise distance determinations. Indeed, the uncertainties 
on iron abundances range from less than 0.1 dex to more than 0.5 dex. 
  
The above empirical scenario is further complicated by the evidence 
that $\omega$ Cen might also be affected by differential reddening 
\citep{dickenscaldwell1988,calamida2005,majewski2012} at a level of 
$\Delta$E(B-V)=0.03-0.04 mag \citep{monibidin2012}.

To overcome the above thorny problems we decided to take advantage of recent 
findings concerning the sensitivity of optical and NIR diagnostics on 
metallicity and reddening to estimate RRL individual distances. 
Pulsation predictions indicate that the spread in magnitude of optical 
and NIR PW relations is smaller when compared with the 
spread typical of optical and NIR PL relations. This finding applies 
to both RRLs and classical Cepheids. The decrease in magnitude dispersion is 
mainly caused by the fact that the PW relations mimic a Period-Luminosity-Color 
(PLC) relation. Thus taking account for the individual position of variable 
stars inside the instability strip 
\citep{bonomarconi1999,udalski1999,soszynski2009,marconi2015}. 
Moreover and even more importantly, theory and observations 
indicate that the PW({\it V\/},\bmv~ and {\it V\/},\bmi) relations display a minimal 
dependence on metallicity. Indeed, their metallicity coefficients are 
at least a factor of two smaller when compared with similar PW relations 
\citep{marconi2015,coppola2015,martinezvazquez2015}.

For the reasons already mentioned in \S~\ref{chapt_empirical_pw_omega} 
(smaller standard deviation, smaller color coefficient) and above, 
we adopted the PW({\it V\/},\bmi) relations to estimate the distance to 
$\omega$ Cen. To quantify possible uncertainties either on the 
zero-point or on the slope, we estimated the distance using the observed 
slope and the predicted zero-point (semi-empirical) and predicted 
PW relation (theoretical, see Table~8 by \citet{marconi2015}).  

Using the metal-independent semi-empirical calibrations 
obtained using the observed slopes and the predicted zero-points  
\citep{marconi2015} we found that the distance modulus to 
$\omega$ Cen (see also Table~\ref{tbl:omegadistance}) ranges from 
13.74$\pm$0.08 (statistical) $\pm$0.01 (systematic) mag (FO) to  
13.69$\pm$0.08$\pm$0.01 mag (F) and to 
13.71$\pm$0.08$\pm$0.01 mag (global). 
The statistical error is the dispersion of the distribution of the distance
moduli of individual RRLs. The systematic error is the difference between the 
theoretical and the semi-empirical calibration of the PW({\it V\/},\bmi) 
relations.
The current estimates agree
within 1$\sigma$ and the mean weighted distance modulus is 
13.71$\pm$0.08$\pm$0.01 mag.
We estimated the distance modulus using also theoretical calibration and we found 
13.74$\pm$0.08$\pm$0.01 mag (FO), 
13.70$\pm$0.08$\pm$0.01 mag (F) and 
13.71$\pm$0.08$\pm$0.01 mag (global). 
The new distance moduli agree with those based on the semi-empirical 
calibration and the mean weighted distance modulus is 13.71$\pm$0.08$\pm$0.01 mag.

The distance moduli that we derived, agree quite well with similar estimates based on the 
{\it K\/}-band PL relation of RRLs provided by 
\citet{longmore1990} (13.61 mag), \citet{sollima2006b} (13.72 mag),  
\citep{bono2008b} (13.75$\pm$0.11 mag),  
by \citet{delprincipe_etal2006} (13.77$\pm$0.07 mag) and 
\citet{navarrete2016} (13.70$\pm$0.03 mag).

A similar remarkable agreement is also found when comparing the current distance 
moduli with those based on the TRGB provided by
\citet{bellazzini2004} (13.70$\pm$0.11 mag) and by \citet{bono2008b} 
(13.65$\pm$0.09 mag). 
The current estimates also agree within 1$\sigma$ with both 
distance moduli provided by \citet{kaluzny_etal2007} 
using cluster eclipsing binaries---namely $\mu$=13.49$\pm$0.14 and 
$\mu$=13.51$\pm$0.12 mag---and with the 
kinematic distance to $\omega$ Cen provided by 
\citet[][$\mu$=13.75$\pm$0.13 mag]{vandeven2006}.
The kinematic distance method applied to GCs is a very promising and independent primary 
distance indicator based on the ratio between the dispersions in proper 
motion and in radial velocity of cluster stars. The key advantage of this 
diagnostic is that its accuracy is only limited by the precision of the 
measurements and by the sample size \citep{kinganderson2002}. The above 
difference seems to suggest the possible unrecognized systematic errors. 
The reader interested in a more detailed discussion concerning the different 
diagnostics adopted to estimate cluster distances is referred to 
\citet{bono2008b}.  

Note that we are not providing independent distance estimates to 
$\omega$ Cen using the zero-point based on the five field RRLs for which are 
available trigonometric parallaxes \citep{benedict2011}. The reason is twofold: 
a) preliminary empirical evidence based on optical, NIR and MIR measurements 
indicates that their individual distances might require a mild revision 
(Neeley et al. 2016, in preparation);  
b) we plan to address on a more quantitative basis the accuracy of 
$\omega$ Cen distance, using optical, NIR and MIR mean magnitudes 
of RRLs (Braga et al. 2016, in preparation).

\section{Metallicity of RR Lyrae stars}\label{chapt_metallicity_omega}

Dating back to the spectroscopic surveys of giant stars by \citet{norris1996} 
and \citet{suntzeffkraft1996}, we have a clear and quantitative
evidence that $\omega$ Cen hosts stellar populations characterized by a 
broad spread in iron abundances. More recently, \citet{fraixburnet2015}, 
by analyzing the abundances provided by \citet{johnson_e_pilachowski2010}, 
confirmed the presence of three main populations as originally suggested by 
\citet{norrisdacosta1995},
\citet{smith2000}, \citet{pancino2002} and \citet{vanture2002}. The general accepted
scenario is that of a globular with a dominant metal-poor primordial population 
(--2.0 $<$ [Fe/H] $<$ --1.6) plus a metal-intermediate (--1.6 $<$ [Fe/H] $<$ --1.3) 
and a relatively metal-rich (--1.3 $<$ [Fe/H] $<$ --0.5) population. The reader 
is also referred to \citet{calamida_etal2009}, for a detailed discussion concerning 
the spread in iron abundance based on the Stroemgren metallicity index for a 
sample of $\sim$4000 stars.  

\begin{figure*}[t]
\centering
\figurenum{15}
\includegraphics[width=13cm]{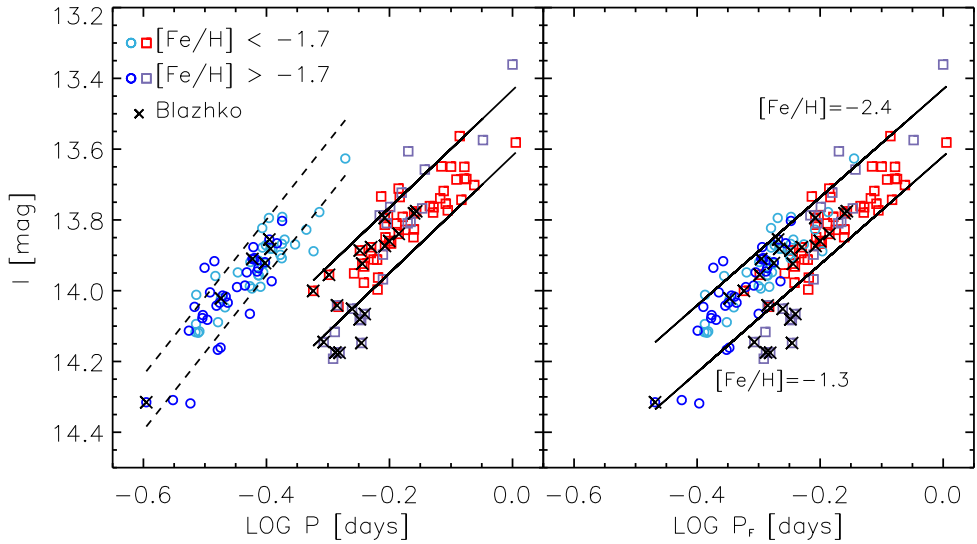}
\vspace*{1.0truecm}
\caption{Left: Empirical {\it I\/}-band PL relation for $\omega$ Cen RRLs. 
Light blue circles and red squares mark variables 
more metal-poor than [Fe/H] = --1.7, while blue circles and violet 
squares mark variables more metal rich than [Fe/H] = --1.7.  
Candidate Blazhko stars are marked with a black cross.
The black lines display predicted \citep{marconi2015} {\it I\/}-band 
PLZ relation for F (solid) and FO pulsators at 
fixed metal abundance  [Fe/H] = --2.4 (brighter) and 
[Fe/H] = --1.3 (fainter).
Right: Same as the left, but for the global (RRc$+$RRab) RRL sample. 
The periods of RRc variables were fundamentalized using 
$\log{P_{RRab}} = \log{P_{RRc}} + 0.127$.} 
\label{fig:plifeh}
\end{figure*}

To further constrain the plausibility of the theoretical framework adopted to 
estimate the distances and to validate the current metallicity scale we compared 
theory and observations in the $\log P$--$I$ plane.   
Figure~\ref{fig:plifeh} shows predicted 
{\it I\/}-band empirical PL relation at different iron abundances (see labeled values)
together with $\omega$ Cen RRLs. Note that 
the objects for which are available iron abundance 
estimates (R00, S06) were plotted using a color 
code: more metal-poor ([Fe/H] $<$ --1.7) 
RRLs are marked with light blue (RRc) and red 
(RRab) colors, while more metal-rich 
([Fe/H] $>$ --1.7) with blue (RRc) and violet (RRab) colors. 
The adopted iron values are based on the R00+S06 homogenized sample, listed 
in column~4 of Table~\ref{tbl:omegametallicity}. The data in this figure 
display two interesting features worth being discussed. 

{\it i)} Predicted PL relation at different iron abundances and observed range 
in iron abundance of RRLs agree quite well, and indeed, the former ones bracket 
the bulk ($\sim$80\%) of the RRL sample. Moreover, there is a mild evidence of a 
ranking in metallicity, indeed more metal-rich RRLs appear---on average---fainter 
than metal-poor ones. This evidence applies to both RRab 
($\Delta I_{poor-rich} \sim 0.15$ mag) and to RRc 
($\Delta I_{poor-rich}\sim 0.09$ mag) variables.

{\it ii)} Blazhko variables are mostly located between RRc and RRab variables. 
Moreover, they also appear to be more associated with more metal-poor (14) 
than with more metal-rich (11) RRLs, the ratio being 1.27. The trend is similar 
to non-Blazhko RRLs, for which the more metal-poor sample (78) is even larger 
than the more metal-rich one (50, the ratio is 1.56). Note that we did not 
take account of RRLs for which iron abundance is not available (35 non-Blazhko
and three Blazhko RRLs).
It is clear that $\omega$ Cen is the right laboratory to 
delineate the topology of the instability strip, due to sample size and the 
broad spread in iron abundance.
Its use is currently hampered by the lack of accurate and precise elemental 
abundances for the entire RRL sample.

On the basis of the above empirical evidence, we decided to take advantage of 
the accuracy of the distance modulus to $\omega$ Cen and of the sensitivity 
of the {\it I\/}-band PL relation to provide a new estimate of the iron abundance 
of individual RRLs. A similar approach was adopted by \citet{martinezvazquez2015} 
and by \citet{coppola2015} to estimate the metallicity distribution of RRLs 
in Sculptor and in Carina, respectively. The absolute {\it I\/}-band magnitudes ($M_I$) 
of RRLs were estimated using the true distance modulus ($\mu$=13.70$\pm$0.02 mag) based on theoretical 
PW relations (see \S~\ref{chapt_distance_omega}). 
In particular, $M_I = I - A_I - \mu$, where $\mu$ is 
the true distance modulus and $A_I$ the selective absorption in the {\it I\/}-band. 
We also adopted, according to \citet{thompson2001,lub2002}, a cluster reddening 
of $E(\bmv)$ = 0.11 mag. We took also account of the spread in $E(\bmv)$ 
measured by \citet{monibidin2012}. 
According to the reddening law provided by 
\citet{cardelli1989}, we adopted a ratio $A_I$/$A_V$ = 0.590. Note that the 
current value accounts for the current photometric system (see for more details 
Section~\ref{chapt_empirical_pw_omega}).

Finally, theoretical {\it I\/}-band PLZ relation for F and FO pulsators 
were inverted to estimate the metallicities of $\omega$ Cen RRLs: 

\begin{equation}\label{eq:invertedplz}
\mathrm{[Fe/H]}=\dfrac{M_I - b\log{P} - a}{c}
\end{equation}

\noindent
where $a$, $b$ and $c$ are the zero-point, the slope and the metallicity
coefficient of the predicted PLZ relations in the form 
$M_I = a+b\log{P}+c\mathrm{[Fe/H]}$. The values of the coefficients 
$a$, $b$ and $c$ are listed in Table~6 of \citet{marconi2015}.
Note that we adopted this relation, 
because theory and observations indicate that PL relations are less prone 
to systematic uncertainties introduced by a spread in stellar mass and/or 
in stellar luminosity due to evolutionary effects \citep{bono2001,bono2003}.
To estimate the iron abundance, we only took into account
RRLs with photometric error in the {\it I\/} band smaller
than 0.1 mag. To provide a homogeneous metallicity 
scale for $\omega$ Cen the above estimates (solar iron abundance in number 
$log{\epsilon_{Fe_{\odot}}}$=7.50, \citealp{pietrinferni2006,marconi2015}) were 
rescaled to the homogeneous cluster metallicity scale provided by 
\citet{gratton2003,carretta2009b} ($log{\epsilon_{Fe_{\odot}}}$=7.54).  

\begin{figure*}[t]
\centering
\figurenum{16}
\includegraphics[width=12cm]{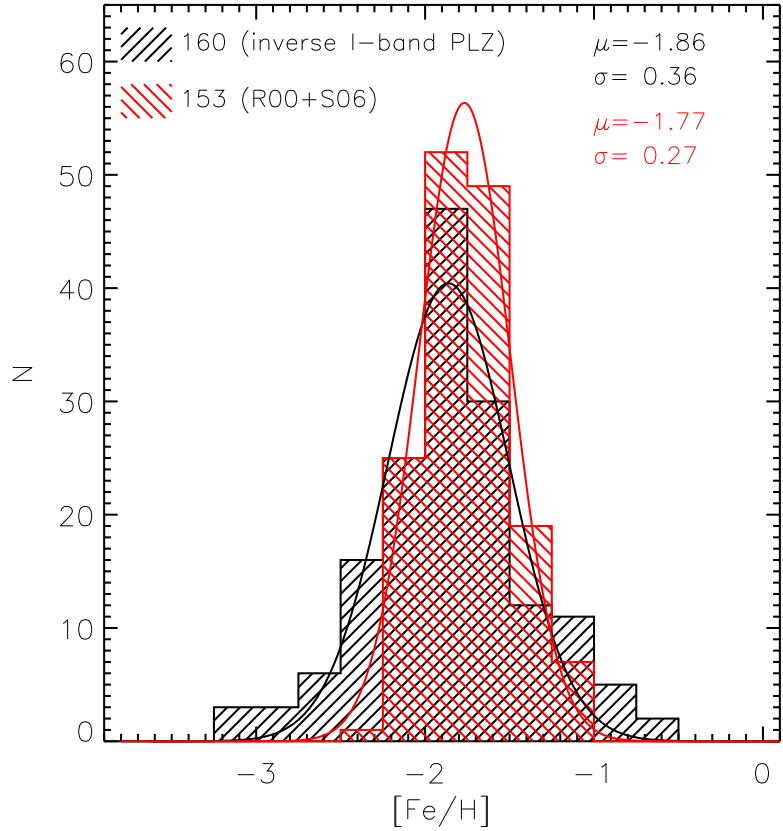}
\vspace*{1.0truecm}
\caption{Iron distribution of $\omega$ Cen RRLs. The black shaded area 
shows iron estimates based on the inversion of theoretical 
{\it I\/}-band PLZ relations for RRab and RRc variables. The red shaded area 
shows iron distribution based on measurements available in the 
literature and based on both spectroscopic  \citep{sollima2006} and 
photometric \citep{rey2000} estimates. The current and the literature 
iron abundances are in the homogeneous cluster metallicity scale provided 
by \citet{carretta2009b}. The red and the black curves 
display the Gaussian fits of the observed distributions. The means and 
the $\sigma$s of the two Gaussians are also labelled.}
\label{fig:histofeh}
\end{figure*}

The metallicity distribution based on 160 RRLs is plotted 
in Fig.~\ref{fig:histofeh} as a black shaded area together with the 
metallicity distribution based on iron abundances provided by R00 and 
by S06 (red shaded area).
We fit the two iron distributions with a Gaussian and we found that 
current distribution is slightly more metal-poor, indeed the difference 
in the peaks is $\Delta$[Fe/H]=0.09. The $\sigma$ of the current distribution 
is larger---0.36 vs 0.27---than the literature one. The difference 
is mainly caused by the fact that the iron distribution based on S06 and R00 
abundances displays a sharp cut-off at [Fe/H]$\sim$--2.3, while the current 
one attains iron abundances that are 0.5 dex more metal-poor.   
We double checked the objects located in the metal-poor tail. Nine out of twelve are 
RRc stars and we found that they mainly belong to the brighter group.
There is also marginal evidence for 
a slightly more extended metal-rich tail, 
but the difference is caused by a few objects.

In passing, we note that the metal-poor tail is only marginally 
supported by both spectroscopic and photometric investigations based on 
cluster red giant stars. Indeed, \citet{calamida_etal2009} using Stroemgren 
photometry for $\sim$4000 red giants, found that the metallicity distribution can 
be fit with seven Gaussians. Their peaks range from [Fe/H]$\sim$--1.7 to [Fe/H]$\sim$+0.2. 
A similar result was also obtained by \citet{johnson_e_pilachowski2010} 
using high-resolution spectra of 855 red giants, suggesting iron 
abundances ranging from [Fe/H] $\sim$ --2.3 to [Fe/H] $\sim$ --0.3 (see also 
\citealp{fraixburnet2015}).

As a consequence of the reasonable agreement in the iron distributions, 
we applied to the current iron distribution the difference in the main 
peaks and provided a homogeneous metallicity scale. For the objects in common 
with R00$+$S06 we computed a mean weighted iron abundance and the final values 
are listed in column~6 of Table~\ref{tbl:omegametallicity}.

It is worth mentioning that the current approach to estimate RRL iron abundances 
depends on the adopted distance modulus. 
A modest increase of 0.05 mag in the true distance modulus implies a 
systematic shift of $\sim$0.30 dex in the peak of the metallicity distribution. 
However, the current approach is aimed at evaluating the relative and not the 
absolute difference in iron abundance. This means that we are mainly interested 
in estimating either the spread (standard deviation) in iron abundance or the 
possible occurrence of multiple peaks in the metallicity distribution 
\citep{martinezvazquez2015}.

\section{Summary and final remarks}\label{chapt_final}

We present new accurate and homogeneous optical, multi-band---{\it UBVRI\/}---photometry 
of the Galactic globular $\omega$ Cen. We collected 8202 CCD images 
that cover a time interval of 24 years and a sky area of 84$\times$48 arcmin across 
the cluster center. The bulk of these images were collected with the Danish  
telescope at ESO La Silla as time-series data in three main long runs 
(more than 4,500 images). The others were collected with several telescopes 
ranging from the 0.9m at CTIO to the VLT at ESO Cerro Paranal. The final 
photometric catalog includes more than 180,000 (Danish) and 665,000 (others) stars 
with at least one measurement in two different photometric bands. The above 
datasets were complemented with optical time series photometry for RRLs 
available in the literature. The global photometric catalog allowed us to 
accomplish the following scientific goals.   

{\em Homogeneity}---We provide new, homogeneous pulsation 
parameters for 187 candidate $\omega$ Cen RRLs. All in all the photometry 
we collected (proprietary$+$literature) covers a time interval of 36 years and the light curves 
of RRLs have a number of phase points per band that ranges from $\sim$10-40 ({\it U\/}), 
to$\sim$ 20-770 ({\it B\/}), to $\sim$20-2830 ({\it V\/}), to $\sim$10-280 ({\it R\/}) 
and to $\sim$10-445 ({\it I\/}).
These numbers sum up to more than 300,000 multi-band phase points for RRLs, 
indicating that this is the largest optical photometric survey ever performed 
for cluster RRLs \citep{jurcsik2012,jurcsik2015}. The above data allowed 
us to provide new and accurate estimates of their pulsation 
parameters (mean magnitudes, luminosity variation amplitudes, epoch of maximum 
and epoch of mean magnitude).

{\em Period distribution}---The key advantage in dealing with 
$\omega$ Cen is that its RRL sample is the 3$^{\mathrm{rd}}$ largest after M3 (237 RRLs) 
and M62 (217) among the globulars hosting RRLs. On the basis of the 
current analysis we ended up with a sample 187 candidate cluster 
RRLs. Among them 101 pulsate in the first overtone (RRc), 85 in the 
fundamental (RRab) mode and a single object is a candidate mixed-mode variable 
(RRd).
We estimate the mean periods for RRab and RRc variables and we found 
that they are $<P_{ab}>= 0.668$ days, $<P_c>= 0.359$ days. The above 
mean periods and the population ratio, i.e., the ratio between 
the number of RRc and the total number of RRLs ($N_c/(N_{ab}+N_d+N_c)$) support 
previous findings suggesting that $\omega$ Cen is a Oosterhoff II cluster.

{\em Bailey Diagram}---The luminosity variation amplitude vs period plane indicates 
a clear lack of HASP RRLs, i.e., RRab 
variables with P$\lesssim$0.48 days and $AV$ $>$ 0.75 mag \citep{fiorentino2015}.
These objects become more popular in stellar systems more metal-rich than 
[Fe/H] $\approx$ --1.4, thus suggesting that RRL in $\omega$ Cen barely 
approach this metallicity range. The RRab variables that, from our investigation, 
appear to be more metal-rich than --1.4, have periods ranging from 0.49 to 0.72 days.

Moreover, we also found evidence that 
RRc can be split into two different groups: 
a) short-period---with periods ranging from $\sim$0.30 to $\sim$0.36 days 
and visual amplitudes ranging from a few hundreths of a magnitude to a few tenths;
b) long-period---with periods ranging from $\sim$0.36 to $\sim$0.45 days
and amplitudes clustering around $AV\sim$0.5 mag. Theoretical and empirical 
arguments further support a well defined spread in iron abundance.

{\em Amplitude ratios}---The well known spread in iron abundance of $\omega$ Cen
stars makes its RRL sample a fundamental test-bench to characterize the possible 
dependence of amplitude ratios on metal content. We performed a detailed test 
and we found that both RRab and RRc attain similar ratios: $AB/AV$ =
1.26$\pm$0.01; $AR/AV$ = 0.78$\pm$0.01; $AI/AV$ = 0.63$\pm$0.01. 
Moreover, they do not display any clear trend with iron abundance.

{\em Visual magnitude distribution}---We performed a detailed analysis of the 
visual magnitude distribution of RRLs and we found that they can be fit with four 
Gaussians. The two main peaks included a significant fraction of RRL ($\sim$76\%) 
and attain similar magnitudes ({\it V\/}$\sim$14.47, 14.56 mag). The fainter  
({\it V\/}$\sim$14.71 mag) and the brighter ({\it V\/}$\sim$14.31 mag) peak include a minor 
fraction (11\%, 13\%) of the RRL sample. The above finding is suggestive of 
a spread in iron abundance of the order of 1.5 dex and paves the way for 
new solid estimates on the absolute age of the different stellar populations 
in $\omega$ Cen.

{\em Blazhko RR Lyrae}---Empirical evidence based on the location of candidate 
Blazhko RRLs in the Bailey diagram and in the color-magnitude diagram 
clearly indicate that they are located between RRc and RRab variables. Indeed, 
we found that a significant fraction (79\%) of them (22 out of 28) have periods
shorter than 0.6 days. Moreover, their location inside the instability strip 
indicates that a significant fraction (39\%) of them belongs to the fainter peak 
(V$\geq$14.6 mag), thus suggesting that this sub-sample is more associated 
with the more metal-rich stellar component.

{\em Oosterhoff dilemma}---Dating back to the seminal investigation by \citet{oosterhoff1939}
in which he recognized that cluster RRLs can be split, according, to their 
mean periods, into two different groups, the astronomical community undertook a 
paramount observational effort in order to constrain the physical mechanism(s) 
driving the empirical evidence. We performed a detailed comparison between 
the period distribution and the Bailey diagram of $\omega$ Cen RRLs with globulars 
hosting a sizable sample (>35) of RRLs and with RRLs in nearby dSphs
and UFDs. We found, as expected, 
that the mean F and FO periods display a steady 
decrease when moving from the more metal-rich (Oosterhoff I) to the more 
metal-poor (Oosterhoff II) clusters. In this context dSphs and UFDs 
attain values that are intermediate between the OoInt and the OoII clusters, 
while $\omega$ Cen appears as the upper envelope of the distribution. On the 
other hand, the population ratio---$N_c/(N_{ab}+N_d+N_c)$---has a nonlinear trend, since 
it attains a well defined minimum for OoInt clusters. In spite of the possible 
differences, the iron abundance appears to be the key parameter in driving the 
transition from short mean periods to long mean periods stellar systems.
The above results do not support the working hypothesis that $\omega$ Cen 
is the core remnant of dwarf galaxy \citep{bekkifreeman2003}. Moreover, 
there is mounting empirical evidence that cluster RRLs might not be the 
appropriate sample to address the Oosterhoff dichotomy, since they  
might be either biased by statistics or affected by environmental 
effects.

{\em $\omega$ Cen disguised as a dwarf galaxy}---The number of globulars hosting long-period 
(0.82--0.85$\lesssim$P$\lesssim$1 days) RRLs is quite limited. Three honorable 
exceptions are $\omega$ Cen and the two metal-rich Bulge globulars hosting
RRLs, namely \ngc{6388} and \ngc{6441}. The mean periods of the metal-rich  
clusters appear as an extreme case of OoI clusters. This is the reason why 
we suggest they should be classified as Oosterhoff type 0 instead of 
Oosterhoff type III. We note that the main common feature among these
clusters is that the HB luminosity function shows a well developed blue tail. 
This indicates that the appearance of long-period RRLs is more nurture that 
nature. The environment, and in particular, the high central stellar density, 
might play a crucial role in the presence of a blue tail, and in turn of 
long-period RRLs. However, the observational scenario appears much more 
complex, since the RRLs in the metal-intermediate cluster \ngc{2808} 
hosts 11 RRab variables, but they have periods shorter than 
0.62 days \citep{kunder2013a}.

{\em Distance determination}---We take advantage of optical PW 
relations that are reddening independent by construction and minimally 
dependent on iron abundance to provide new estimates of the distance 
to $\omega$ Cen. We adopted both a semi-empirical and a theoretical 
calibration and we found a true distance modulus of 13.71$\pm$0.08$\pm$0.01 mag. 
They agree quite well with similar estimates available in 
the literature. In particular, we found that the agreement is within 1$\sigma$  
with the geometrical distances based on eclipsing binaries 
\citep[13.49$\pm$0.14 / 13.51$\pm$0.12,][]{kaluzny_etal2007}.

{\em Metallicity distribution}---We inverted the {\it I\/}-band 
PLZ relation for F and FO pulsators 
to provide individual metallicity estimates for 160 cluster RRLs. 
We found that the metallicity distribution agrees quite well with the 
metallicity distribution of RRLs based on spectroscopic measurements 
(74, S06) and on photometric indicators (131, R00). We also found 
evidence of a metal-poor tail that is not present in previous 
spectroscopic investigations of $\omega$ Cen RRLs.    

The current long-term photometric surveys are providing 
new and homogeneous measurements concerning field and cluster stars. The current 
status is going to experience a quantum jump as soon as the ongoing (Gaia) and 
near future ground-based experiments will release their data. The project we 
started more than 15 years ago on $\omega$ Cen, may be defined as a {\em local survey}.     
The number of optical images adopted to individuate main sequence and evolved 
variable stars have been discussed in detail on
Section~\ref{chapt_obs_opt}. 
In dealing with the
Danish dataset we analyzed 4539 images and we performed $\approx2.5\times10^8$ measurements,
which means roughly 55,000 stars per image. In dealing with all the other
optical datasets we analyzed 3663 images and performed $\approx1.1\times10^8$ measurements,
which means roughly 27,000 stars per image. A similar number of measurements have 
been also performed in dealing with NIR images. The above numbers indicate once accounting 
for the preliminary steps in approaching the final photometric catalog, that we are 
dealing with an experiment that included more than one giga measurements. 
The results concerning variable and static stars will be addressed in a series
of future papers in which we plan to use homogeneous multi-band optical, NIR 
and MIR photometry.

\acknowledgements 
It is a pleasure to thank the many colleagues that during the last 15 years with 
their support and contribution made this experiment possible. One of us (V.F.B.) 
thanks ESO for a science visit during which part of this paper was written, while 
G.B. thanks the Japan Society for the Promotion of Science
for a research grant (L15518) and G.F. thanks the support from FIRB~2013 
(grant: RBFR13J716). 
This research has made use of the USNO Image and Catalogue Archive
operated by the United States Naval Observatory, Flagstaff Station
(\url{http://www.nofs.navy.mil/data/fchpix/}). This research has made 
use of NASA's Astrophysics Data System.

\appendix 
\section{Notes on individual RR Lyrae stars} 

{\it V4, V25, V44, V88, V90, V91, V271, V272, V273, V276, 
NV340, NV341, NV349, NV350, NV352---} We estimate the pulsation parameters 
neglecting either the {\it danish95} and/or the  {\it danish98} and/or the 
{\it danish99} datasets, since they are noisy.\\

{\it V5, V9, V11, V56, V67, V69, V74, V106, V112, V115, V120, V130, V140, V141---} 
We confirm the Blazhko modulations suggested by 
\citet{kaluzny_etal2004} and \citet{weldrake_etal2007} and provide 
preliminary estimates of {\it B\/}- and/or 
{\it V\/}-band Blazhko amplitudes (see Table~\ref{tab:phot_rr_opt}).\\

{\it V10, V24, V32, V47, V58, V64, V70, V71, V77, V81, V82, V87, V89, V95, 
V123, V124, V126, V131, V136, V145, V147, V153, V155, V156, V157, V158, V166, V270, V275, V289,
NV340, NV346, NV347, NV353, NV354---} There is mild evidence of a period change.\\

{\it V11, V94---} There is evidence that these variables might be affected by 
a phase shift. Owing to these variations, the {\it B\/}- and {\it V\/}-band mean magnitudes 
and amplitudes are based either on the {\it danish95} or on the OGLE dataset.\\

{\it V22, V30, V32, V94, V261, V275, V280, V291---} The current 
photometry suggests that these objects are new candidate Blazhko variables.
The current data do not allow us to support the multi-modality for V261, 
suggested by \citet{kaluzny_etal2004} and/or possible variations in the 
pulsation period. The variables V280 and V291 sow also evidence for a 
secondary modulations and/or for a phase shift.\\

{\it V45, V165---} Candidate Blazhko variable according to \citet{kaluzny_etal2004}. 
The current light curves, based only on the {\it other} dataset, 
are poorly sampled and do not allow us to deduce the 
occurrence of a Blazhko modulation.\\

{\it V52---} The pulsation parameters are based on the {\it other} dataset, since 
the variable is blended in all Danish datasets. Moreover, according to 
\citet{navarrete_etal2015}, a neighboring star is located at $\sim$0.5\farcs from the RRL variable.\\

{\it V55---} Period, mean magnitudes and amplitudes are based on the photometry by J. Lub and \citet{sturch1978}.\\

{\it V59, V82, V97---} Candidate Blazhko variables according to \citet{kaluzny_etal2004}. 
The current light curves are well sampled, but they do not show evidence of Blazhko modulation.  
No firm conclusion can be reached on their Blazhko nature.\\

{\it V68---} This is the brightest ({\it V\/}$\sim$14.24 mag) and the longest period (0.53476174 days) 
RRc variable. This is an interesting object worth being investigated in more detail.\\

{\it V73---} Candidate Blazhko variable according to \citet{martin1938}. 
The current light curves are only based on the {\it other} dataset. They are 
poorly sampled and do not allow us to deduce the occurrence of a Blazhko 
modulation.\\

{\it V80, V177, NV411, NV433---} These stars lie outside the area covered by our images. We only provide a
new estimate of the periods from the {\it V\/}+{\it R\/} band light curve by
\citet{weldrake_etal2007}. NV433 is also a candidate field variable and the current data
do not allow us to address whether it is an RRL \citep{weldrake_etal2007,navarrete_etal2015}.\\

{\it V84---} This star lies outside the area covered by our images. The pulsation 
parameters are based on Walraven {\it BV\/}-band photoelectric photometry 
performed by J. Lub, on {\it UBV\/} photoelectric photometry provided by 
\citet{sturch1978} and on OGLE {\it V\/}-band photometry \citep{kaluzny_etal1997}. \\

{\it V142---} Preliminary results concerning the mode identification of V142 indicate that it is the 
first radial double-mode pulsator in $\omega$ Cen. Note that the double-mode variables
found by \citet{olech2009} are not F+FO pulsators.\\

{\it V168---} This RRL is a candidate field variable \citep{vanleeuwen_etal2000,bellini2009}. 
On the basis of the current {\it V\/}-band mean magnitude we confirm its non-membership. There is 
mild evidence of a period change.\\

{\it V172, NV457, NV458---} These stars lie outside the area covered by our images. The period, mean magnitude,
amplitude and epochs of maximum and minimum light have been derived from {\it V\/} photometry
by the CATALINA survey \citep{drake_etal2009,drake13a,drake13b,drake14,torrealba15}.\\

{\it V181, V183---} These stars lie outside the area covered by our images. 
On the basis of their position in the {\it K\/},\jmk CMD, these were
classified as a candidate field variable stars \citep{navarrete_etal2015}.\\

{\it V263, NV366---} These variables have periods of 1.01215500 and 0.99992364 days 
and are located on the transition between RRLs and TIICs. The RRL--TIIC transition
will be discussed in a forthcoming paper.\\

{\it V281, V283---} These stars lie outside the area covered by our images. The periods, mean magnitudes,
amplitudes and epochs of maximum and minimum light have been derived from {\it V\/} photometry
by OGLE \citep{kaluzny_etal1997}. V283 is also a candidate field variable.\\

{\it NV351---} This variable on our images is heavily blended and we could not derive the mean magnitudes and 
amplitudes.\\

\bibliographystyle{apj}

\clearpage
\tablenum{1}
\begin{deluxetable*}{llclll rrrrrr c}
\tablewidth{0pt}
\tabletypesize{\scriptsize}
\tablecaption{Log of the observations of $\omega$ Cen in optical bands.}
\tablehead{Run & ID & P/A\tablenotemark{a} & Dates & Telescope & Camera & {\it U\/} & {\it B\/}& {\it V\/}& {\it R\/}& {\it I\/}& other & multiplex \\}
\startdata
     1 & bond24    & A &  1988 Feb 22--Mar 02        & CTIO 0.9m       & RCA5                 &  \ldots &    3 &    3 &   3 &   3 &   \ldots &    \\
     2 & bond23    & A &  1989 Jan 16--25            & CTIO 0.9m       & RCA5                 &  \ldots &    1 &    1 &   1 &   1 &   \ldots &    \\
     3 & f32       & P &  1991 Jan 22                & CTIO 1.5m       & Tek 512              &  \ldots &    3 &    3 &   \ldots &   \ldots &   \ldots &    \\
     4 & emmi6     & A &  1993 Feb 17--21            & ESO NTT 3.6m    & EMMI susi3?          &  \ldots &    \ldots &   47 &   \ldots &  47 &   \ldots &    \\
     5 & emmi7     & A &  1993 Feb 18--21            & ESO NTT 3.6m    & EMMI susi3?          &  \ldots &    \ldots &    2 &   \ldots &   2 &   \ldots &    \\
     6 & emmi5     & A &  1993 Jul 15--23            & ESO NTT 3.6m    & EMMI                 &  \ldots &    \ldots &    2 &   \ldots &   2 &   \ldots &    \\
     7 & susi1     & A &  1994 Dec 26--29            & ESO NTT 3.6m    & SUSI                 &  \ldots &    2 &    3 &   \ldots &   4 &   \ldots &    \\
     8 & emmi2     & A &  1995 Mar 07--10            & ESO NTT 3.6m    & EMMI                 &  \ldots &    \ldots &    7 &   \ldots &   7 &   \ldots &    \\
\enddata
\end{deluxetable*}

\tablenum{2}
\begin{deluxetable*}{l c c c c cccc}
\tablewidth{0pt}
\tabletypesize{\footnotesize}
\tablecaption{Positions and periods for $\omega$ Cen RRLs.}
\tablehead{ID & 
\multicolumn{1}{c}{$\alpha$ (J2000.0)\tablenotemark{a}}   & 
\multicolumn{1}{c}{$\delta$ (J2000.0)\tablenotemark{a}}   & 
\multicolumn{2}{c}{Period\tablenotemark{b}} & 
\multicolumn{2}{c}{$T_0(max)$\tablenotemark{c}} &
\multicolumn{2}{c}{$T_0(rising)$\tablenotemark{c}} \\
        & h \quad m \quad s  & \deg \quad \farcm \quad \farcs & Literature\tablenotemark{d} & LS & band & JD & band & JD }
\startdata
V3                        &  13  25  56.16  &  --47  25  54.2 &  0.8413  &    0.841262  & \ldots &     \ldots   & \ldots &     \ldots   \\ 
V4                        &  13  26  12.94  &  --47  24  19.2 &  0.6273  &    0.627318  & \ldots &      \ldots  &      V &  49156.5732 \\  
V5                        &  13  26  18.34  &  --47  23  12.8 &  0.5153  &    0.515280  &      V &   50975.5712 &      V &  49865.6237 \\  
V7                        &  13  27  01.04  &  --47  14  00.1 &  0.7130  &    0.713034  &      V &   49142.5355 &      V &  49082.5766 \\  
V8                        &  13  27  48.43  &  --47  28  20.6 &  0.5213  &    0.521326  &      V &   51285.8083 &      V &  49824.5018 \\  
V9                        &  13  25  59.59  &  --47  26  24.4 &  0.5235  &    0.523464  & \ldots &      \ldots  & \ldots &     \ldots   \\ 
V10                       &  13  26  07.01  &  --47  24  37.0 &  0.3750  &    0.374882  &      I &   49863.6418 &      I &  52446.5061 \\  
V11                       &  13  26  30.56  &  --47  23  01.9 &  0.5648  &    0.564806  & \ldots &      \ldots  & \ldots &     \ldots  \\
\enddata                                                                                 
\tablenotetext{a}{~The mean epoch associated to these coordinates is May 2004.}
\tablenotetext{b}{~Pulsation periods (days) based either on a compilation of literature values \citep{clement2001} or 
on the current estimate using the LS method.}
\tablenotetext{c}{~Epoch of light maximum and of mean magnitude on the rising
branch (HJD--2,400,000) estimated using the spline fit to optical light curves.} 
\tablenotetext{d}{~V80, V172 and V173 from ~\citet{navarrete_etal2015}, V175 from \citet{fernandeztrincado2015}, other from \citet{clement2001}.}
\tablenotetext{e}{~Coordinates from the Clement's catalog.}
\tablenotetext{f}{~Candidate field RRL.}
\tablenotetext{g}{~Coordinates from \citet{fernandeztrincado2015}.}
\tablenotetext{h}{~Coordinates from \citet{navarrete_etal2015}.}
\label{tab:pos_rr}
\end{deluxetable*}

\tablenum{3}
\setlength{\tabcolsep}{0.1cm}
\begin{deluxetable*}{l ccccc ccccc c l}
\tablewidth{0pt}
\tabletypesize{\scriptsize}
\tablecaption{Optical---{\it UBVRI\/}---mean magnitudes and amplitudes for $\omega$ Cen RRLs.}
\tablehead{
ID & 
{\it U\/}\tablenotemark{a}& 
{\it B\/}\tablenotemark{a}& 
{\it V\/}\tablenotemark{a}& 
{\it R\/}\tablenotemark{a}& 
{\it I\/}\tablenotemark{a}& 
$AU$\tablenotemark{b} & 
$AB$\tablenotemark{b} & 
$AV$\tablenotemark{b} & 
$AR$\tablenotemark{b} & 
$AI$\tablenotemark{b} & 
$q_{UBVRI}$\tablenotemark{c} & 
Mode\tablenotemark{d} \\
                       & mag & mag  & mag  & mag & mag & mag & mag & mag & mag  & mag & & }
\startdata                   
{      V3 }&{ 15.017$\pm$0.015 }&{ 14.888$\pm$0.005 }&{ 14.391$\pm$0.004 }&{     \ldots       }&{ 13.685$\pm$0.010 }&{ 1.054 }&{ 0.934            }&{ 0.761          }&{\ldots }&{ 0.401 }&{ 12301 }&{  RRab} \\
{      V4 }&{ 15.207$\pm$0.185 }&{ 14.882$\pm$0.005 }&{ 14.467$\pm$0.009 }&{ 14.462$\pm$0.180 }&{ 13.787$\pm$0.018 }&{ \ldots}&{ 1.525            }&{ 1.119          }&{\ldots }&{ 0.904 }&{ 23322 }&{  RRab} \\
{      V5 }&{ 15.303$\pm$0.028 }&{ 15.197$\pm$0.013 }&{ 14.702$\pm$0.011 }&{ 14.359$\pm$0.012 }&{ 14.174$\pm$0.031 }&{ 0.972 }&{ 1.021            }&{ 0.852-1.271    }&{ 1.118 }&{ 0.437 }&{ 13332 }&{ RRab*} \\
{      V7 }&{ 15.068$\pm$0.017 }&{ 14.896$\pm$0.010 }&{ 14.594$\pm$0.010 }&{     \ldots       }&{ 13.768$\pm$0.012 }&{ 0.739 }&{ 0.885            }&{ 0.950          }&{\ldots }&{ 0.293 }&{ 11301 }&{  RRab} \\
{      V8 }&{ 15.126$\pm$0.022 }&{ 15.016$\pm$0.008 }&{ 14.671$\pm$0.005 }&{     \ldots       }&{ 14.046$\pm$0.014 }&{ 1.368 }&{ 1.467            }&{ 1.263          }&{\ldots }&{ 0.695 }&{ 11302 }&{  RRab} \\
{      V9 }&{ 15.592$\pm$0.403 }&{ 15.204$\pm$0.010 }&{ 14.779$\pm$0.006 }&{ 14.498$\pm$0.179 }&{ 14.175$\pm$0.023 }&{ \ldots}&{ 1.348            }&{ 0.700-1.170    }&{\ldots }&{ 0.516 }&{ 22322 }&{ RRab*} \\
{     V10 }&{ 14.869$\pm$0.031 }&{ 14.826$\pm$0.017 }&{ 14.505$\pm$0.013 }&{ 14.027$\pm$0.018 }&{ 13.919$\pm$0.024 }&{ 0.694 }&{ 0.565            }&{ 0.421          }&{ 0.539 }&{ 0.280 }&{ 12323 }&{   RRc} \\
{     V11 }&{ 15.045$\pm$0.020 }&{ 14.866$\pm$0.026 }&{ 14.476$\pm$0.035 }&{ 14.406$\pm$0.221 }&{ 13.886$\pm$0.032 }&{ 0.675 }&{ 0.784-1.261      }&{ 0.453-1.017    }&{\ldots }&{ 0.576 }&{ 22323 }&{ RRab*} \\
\enddata
\tablenotetext{a}{Mean {\it UBVRI\/} magnitudes. The means were computed as intensity averages 
and then transformed into magnitudes.} 
\tablenotetext{b}{Luminosity amplitudes in {\it UBVRI\/}. The amplitudes were estimated as 
the difference between the minimum and the maximum of the spline fit. The {\it U\/}-band 
amplitude is only available for a few variables (see Appendix). The minimum 
and maximum amplitudes of candidate Blazhko RRLs and of the candidate mixed-mode 
variable were estimated as the amplitudes of the lower and of the upper envelope 
of the observed data points.}
\tablenotetext{c}{Photometric quality index of the light curves taking account 
of the phase coverage, uncertainties on individual measurements and photometric 
scatter. The ranking is: 0, not available; 1, poor; 2, decent; 3, good.}
\tablenotetext{d}{Pulsation mode identification. RRab: fundamental; 
RRc: first overtone; RRd: mixed-mode . An asterisk marks the variables that are 
candidate Blazhko RRLs.}
\label{tab:phot_rr_opt}
\end{deluxetable*}

\tablenum{4}
\begin{deluxetable*}{l c c c c c c}
\tablewidth{0pt}
\tabletypesize{\scriptsize}
\tablecaption{Mean amplitude ratios for RRab, RRc and for the global (All) sample of 
$\omega$ Cen RRLs.}
\tablehead{   
              & \multicolumn{2}{c}{RRab} & \multicolumn{2}{c}{RRc} & \multicolumn{2}{c}{All}  \\
              &   mean    &    $\sigma$  &   mean    &  $\sigma$ &   mean    &  $\sigma$       }
\startdata                      
$AB$/$AV$ & 1.25$\pm$0.01 & 0.11 & 1.26$\pm$0.02 & 0.16 & 1.26$\pm$0.01 & 0.12 \\
$AR$/$AV$ & 0.80$\pm$0.03 & 0.10 & 0.77$\pm$0.02 & 0.12 & 0.78$\pm$0.01 & 0.11 \\
$AI$/$AV$ & 0.63$\pm$0.01 & 0.07 & 0.63$\pm$0.01 & 0.10 & 0.63$\pm$0.01 & 0.08 \\
$AV$/$AI$ & 1.57$\pm$0.02 & 0.13 & 1.63$\pm$0.03 & 0.23 & 1.60$\pm$0.02 & 0.20 \\
$AB$/$AR$ & 1.59$\pm$0.07 & 0.28 & 1.57$\pm$0.04 & 0.23 & 1.57$\pm$0.03 & 0.24 \\
$AB$/$AI$ & 1.96$\pm$0.02 & 0.14 & 2.03$\pm$0.04 & 0.32 & 2.00$\pm$0.02 & 0.26 
\enddata
\label{tab:amplratio}
\end{deluxetable*}

\tablenum{5}
\begin{deluxetable*}{l c c r c}
\tablewidth{0pt}
\tabletypesize{\scriptsize}
\tablecaption{Parameters of the four Gaussians adopted to fit the {\it V\/}-band magnitude 
distribution of $\omega$ Cen RRLs.}
\tablehead{ Fit\tablenotemark{a}  &  $\mu$  & $\sigma$ & $A$  & Relative area }
\startdata                     
1 & 14.47 &  0.06  &  29.95  &  0.51 \\
2 & 14.56 &  0.07  &  12.62  &  0.25 \\
3 & 14.71 &  0.08  &   4.81  &  0.11 \\
4 & 14.31 &  0.11  &   4.01  &  0.13 
\enddata
\tablenotetext{a}{Gaussian fit of the form A*$\exp{((x-x_0)/\sigma)^2}$.}
\label{tab:gaussfit}
\end{deluxetable*}

\tablenum{6}
\begin{deluxetable*}{l c c c c c c}
\tablewidth{0pt}
\tabletypesize{\scriptsize}
\tablecaption{Pulsation properties of RRLs in different gas poor stellar systems and 
in $\omega$ Cen.}
\tablehead{ System & <$P_{c}$>\tablenotemark{a} & $N_c$ & <$P_{ab}$>\tablenotemark{a} & $N_{ab}$ & $N_c$/$N_{tot}$\tablenotemark{b} & [Fe/H]\tablenotemark{c} \\
                   &   days   &  & days  &    &                 & }
                   \startdata                  
\multicolumn{7}{c}{OoIII/Oo0} \\
\ngc{6388} &           0.384$\pm$0.017 &   18 & 0.739$\pm$0.037 &   13 & 0.58  & --0.55 \\
\ngc{6441} &           0.380$\pm$0.016 &   23 & 0.755$\pm$0.016 &   47 & 0.32  & --0.45 \\
\hline
Total     &         0.382$\pm$0.012 &   41 & 0.752$\pm$0.015 &   60 & 0.40  & --0.55/--0.45 \\ 
\hline
\hline
\multicolumn{7}{c}{OoI} \\
\ngc{5272}        &    0.328$\pm$0.007 &   46 & 0.564$\pm$0.005 &  135 & 0.25  &  --1.50 \\
\ngc{5904}        &    0.318$\pm$0.005 &   38 & 0.551$\pm$0.008 &   86 & 0.31  &  --1.29 \\
\ngc{6121}        &    0.293$\pm$0.011 &   14 & 0.548$\pm$0.014 &   31 & 0.31  &  --1.16 \\
\enddata
\tablenotetext{a}{Mean FO and F period (days).}  
\tablenotetext{b}{Ratio between the number of FO ($N_c$) and the total ($N_{ab}$+$N_d$+$N_c$) number of RRLs.}  
\tablenotetext{c}{Range in iron abundance covered by the selected stellar systems. 
Iron abundances for GCs come from the \citep{harris1996} catalog, while dSphs and UFDs 
from \citet{kirby2013}; \citet{mcconnachie2012}; \citet{fabrizio2015}.}
\label{tab:oosterhoff_comparison}
\end{deluxetable*}

\tablenum{7}
\begin{deluxetable*}{rrc rrc rrc}
\tablewidth{0pt}
\tabletypesize{\scriptsize}
\tablecaption{Empirical {\it I\/}-band Period-Luminosity relations of the form 
$M_I$=$a$ + $b\log{P}$ for $\omega$ Cen RRLs.}
\tablehead{
a{\tablenotemark{a}}& b{\tablenotemark{a}}& $\sigma${\tablenotemark{a}}& a{\tablenotemark{b}}&
b{\tablenotemark{b}}& $\sigma${\tablenotemark{b}}& a{\tablenotemark{c}} & b{\tablenotemark{c}} & $\sigma${\tablenotemark{c}} \\
                       mag & mag & mag & mag & mag  & mag  & mag & mag & mag}
\startdata
     \multicolumn{3}{c}{RRc} & \multicolumn{3}{c}{RRab} & \multicolumn{3}{c}{All} \\
  13.252 &      --1.624 &  0.058 &      13.485 &      --1.955 &  0.078 &      13.563 &      --1.335 &  0.061 \\
$\pm$ 0.047 & $\pm$ 0.105 &       & $\pm$ 0.026 & $\pm$ 0.129 &       & $\pm$ 0.015 & $\pm$ 0.053 &      
     \enddata
\tablenotetext{a}{Zero-point (a), slope (b) and standard deviation ($\sigma$) for first overtone (RRc) variables.
The errors on the zero-point and on the slope are listed in the 2nd row.}
\tablenotetext{b}{Zero-point (a), slope (b) and standard deviation ($\sigma$) for fundamental (RRab) variables.
The errors on the zero-point and on the slope are listed in the 2nd row.}
\tablenotetext{c}{Zero-point (a), slope (b) and standard deviation ($\sigma$) for the global RRL sample (All). 
The periods of RRc variables were fundamentalized by adopting $\log{P_{F}} = \log{P_{FO}} + 0.127$.
The errors on the zero-point and on the slope are listed in the 2nd row.}
\label{tbl:omegacen_pl_empirical}
\end{deluxetable*}

\tablenum{8}
\begin{deluxetable*}{lc rrc rrc rrc}
\tablewidth{0pt}
\tabletypesize{\scriptsize}
\tablecaption{Empirical optical Period-Wesenheit relations for $\omega$ Cen RRLs.}
\tablehead{
PW{\tablenotemark{a}} &  $\zeta$\tablenotemark{b} & a{\tablenotemark{c}}& b{\tablenotemark{c}}& 
$\sigma${\tablenotemark{c}}& a{\tablenotemark{d}}&
b{\tablenotemark{d}}& $\sigma${\tablenotemark{d}}& a{\tablenotemark{e}} & b{\tablenotemark{e}} & $\sigma${\tablenotemark{e}} \\
                     &  & mag & mag & mag & mag & mag  & mag  & mag & mag & mag}
\startdata
       &   & \multicolumn{3}{c}{RRc} & \multicolumn{3}{c}{RRab} & \multicolumn{3}{c}{All} \\
\multicolumn{10}{c}{Dual-band PW relations} \\ 
{\it V\/},\bmv  & 3.06 & 12.242 &      -2.750 &  0.136 &      12.622 &      -2.919 &  0.122 &      12.670 &      -2.574 &  0.132 \\
              &   & $\pm$ 0.110 & $\pm$ 0.243 &       & $\pm$  0.041 & $\pm$ 0.204 &       & $\pm$  0.031 & $\pm$ 0.112 & \\
{\it I\/},\bmi  & 0.78 & 12.070 &      -2.734 &  0.081 &      12.460 &      -2.868 &  0.075 &      12.511 &      -2.503 &  0.079 \\
              &   & $\pm$ 0.065 & $\pm$ 0.144 &       & $\pm$  0.025 & $\pm$ 0.124 &       & $\pm$  0.019 & $\pm$ 0.068 & \\
{\it I\/},\vmi  & 1.38 & 11.990 &      -2.765 &  0.093 &      12.442 &      -2.658 &  0.086 &      12.460 &      -2.443 &  0.089 \\
              &   & $\pm$ 0.074 & $\pm$ 0.165 &       & $\pm$  0.030 & $\pm$ 0.149 &       & $\pm$  0.021 & $\pm$ 0.077 & \\
\enddata
\tablenotetext{a}{PW relations of the form: $W(X,Y-Z)$ = a + b$\log\ P$, where
$W(X,Y-Z) = X + (A_X/(A_Y-A_Z))(Y-Z)$ is the Wesenheit magnitude. {\it Z\/}$\neq${\it X\/} only for triple-band relations.}
\tablenotetext{b}{Color coefficient in Wesenheit magnitude: $\zeta$=$A_X/(A_Y-A_Z)$}
\tablenotetext{c}{Zero-point (a), slope (b) and standard deviation ($\sigma$) for first overtone (RRc) variables.
The errors on the zero-point and on the slope are listed in the 2nd row.}
\tablenotetext{d}{Zero-point (a), slope (b) and standard deviation ($\sigma$) for fundamental (RRab) variables.
The errors on the zero-point and on the slope are listed in the 2nd row.}
\tablenotetext{e}{Zero-point (a), slope (b) and standard deviation ($\sigma$) for for the global RRL sample (All). 
The periods of RRc variables were fundamentalized by adopting $\log{P_{F}} = \log{P_{FO}} + 0.127$.
The errors on the zero-point and on the slope are listed in the 2nd row.}
\label{tbl:omegacen_pw_empirical}
\end{deluxetable*}

\clearpage
\tablenum{9}
\begin{deluxetable*}{l l l c}
\tablewidth{0pt}
\tabletypesize{\scriptsize}
\tablecaption{True distance moduli to $\omega$ Cen: literature and current estimates.}
\tablehead{$\mu_0$ & E(\bmv)\tablenotemark{a} & Ref. & Notes\tablenotemark{b} \\ mag & mag & & }
\startdata  
13.61                           & 0.11          & \citet{longmore1990}   & (1)  \\
13.65$\pm$0.12                  & 0.13          & \citet{thompson2001} & (2)  \\
13.70$\pm$0.11                  & 0.11$\pm$0.01 & \citet{bellazzini2004} & (3)  \\
13.77$\pm$0.07                  & 0.11$\pm$0.01 & \citet{delprincipe_etal2006} &   (4)  \\
13.72                           & 0.11$\pm$0.01 & \citet{sollima2006b} &   (5)  \\
13.75$\pm$0.13                  & 0.11          & \citet{vandeven2006} &   (6)  \\
13.49$\pm$0.14~/~13.51$\pm$0.12 & 0.131         & \citet{kaluzny_etal2007} & (7) \\
13.65$\pm$0.09                  & 0.11$\pm$0.02 & \citet{bono2008b} & (8) \\
\enddata
\end{deluxetable*}

\clearpage
\tablenum{10}
\begin{deluxetable*}{l c c c c c}
\tablewidth{0pt}
\tabletypesize{\scriptsize}
\tablecaption{Metallicity estimates ([Fe/H]) of candidate $\omega$ Cen RRLs based on 
spectroscopy (S06) and on photometric indices (R00, our).}
\tablehead{ 
& \multicolumn{5}{c}{[Fe/H]} \\
ID & S06\tablenotemark{a}   & Rey\tablenotemark{b}   & S06$+$Rey\tablenotemark{c} & PLI\tablenotemark{d} & S06$+$Rey$+$PLI\tablenotemark{e}}
\startdata   
 V3            &        \ldots     & --1.54$\pm$0.05  &  --1.73$\pm$0.05 &   --1.59       &   --1.72$\pm$0.04 \\                                                              
 V4            &        \ldots     & --1.74$\pm$0.05  &  --1.95$\pm$0.05 &   --2.24       &   --1.96$\pm$0.08 \\                                                    
 V5            &   --1.24$\pm$0.11 & --1.35$\pm$0.08  &  --1.43$\pm$0.17 &   --0.78       &   --1.22$\pm$0.43 \\                                                    
 V7            &        \ldots     & --1.46$\pm$0.08  &  --1.64$\pm$0.08 &   --1.80       &   --1.66$\pm$0.07 \\                                                    
 V8            &        \ldots     & --1.91$\pm$0.28  &  --2.13$\pm$0.28 &   --1.49       &   --1.77$\pm$0.45 \\                                                    
 V9            &        \ldots     & --1.49$\pm$0.06  &  --1.67$\pm$0.06 &   --0.70       &   --1.62$\pm$0.31 \\                                                    
 V10           &        \ldots     & --1.66$\pm$0.10  &  --1.86$\pm$0.10 &   --1.90       &   --1.87$\pm$0.02 \\                                                    
 V11           &   --1.61$\pm$0.22 & --1.67$\pm$0.13  &  --1.81$\pm$0.15 &   --2.09       &   --1.89$\pm$0.18 \\
\enddata
\end{deluxetable*}


\clearpage

\end{document}